\documentclass[]{interact}

\usepackage{fix-cm}

\usepackage[numbers,sort&compress]{natbib}
\bibpunct[, ]{[}{]}{,}{n}{,}{,}
\renewcommand\bibfont{\fontsize{10}{12}\selectfont}
\makeatletter
\def\NAT@def@citea{\def\@citea{\NAT@separator}}
\makeatother

\usepackage{epstopdf}       
\usepackage[caption=false]{subfig}
\usepackage[onehalfspacing]{setspace}
\usepackage{etoolbox}
\usepackage{amsmath,amssymb}
\usepackage{algorithm}
\usepackage{algorithmic}
\usepackage{caption}
\usepackage{url,hyperref,microtype,tikz,pgfplots,float}
\usepackage{xcolor}  
\usepackage{adjustbox}  
\pgfplotsset{compat=newest}

\newcommand{\raj}[1]{\textcolor{black}{#1}}  

\usepackage{amsthm}
\theoremstyle{plain}

\theoremstyle{definition}

\theoremstyle{remark}

\title{EvoSort: A Genetic-Algorithm-Based Adaptive Parallel Sorting Framework for Large-Scale High Performance Computing}

\author{
\name{Shashank Raj\textsuperscript{a,*}\thanks{CONTACT Shashank Raj. Email: \texttt{rajshash@msu.edu}} 
and Kalyanmoy Deb\textsuperscript{b,*}\thanks{CONTACT Kalyanmoy Deb. Email: \texttt{kdeb@msu.edu}}}
\affil{\textsuperscript{a}Department of Computer Science and Engineering, 
Michigan State University, East Lansing, MI, USA;\\
\textsuperscript{b}Department of Electrical and Computer Engineering, 
Michigan State University, East Lansing, MI, USA}
}

\begin{document}
\maketitle

\begin{abstract} \raj{We present EvoSort, a general-purpose adaptive parallel sorting framework accessible at the Python level. EvoSort employs a Genetic Algorithm (GA) to automatically discover and refine critical parameters, including insertion sort thresholds and algorithm selection (mergesort vs. LSD radix sort). By adapting continuously to input data and system architecture, EvoSort provides a drop-in replacement for standard Python routines like NumPy and Pandas. Experiments on up to $10^{10}$ (10 billion) elements across nine data distributions and two hardware platforms demonstrate that EvoSort consistently outperforms competing methods. Results show speedups of up to $225\times$, exemplifying a powerful auto-tuning solution for large-scale data processing.}

\vspace{1ex} \tiny \textbf{Keywords:} High-Performance Computing, Parallel Sorting, Genetic Algorithm, Auto-tuning, Radix Sort, Mergesort \end{abstract}

\section{Introduction}

\raj{Sorting is a fundamental operation that underpins a wide range of computational tasks in science, engineering, commerce and data analytics \citep{Knuth1998,Cormen2009}. Whether it is used in database management, scientific simulations, or real-time data processing, the efficiency of sorting algorithms has a profound impact on overall system performance \citep{Sanders2004,Shun2015}. In today's era of big data and high-performance computing (HPC), datasets frequently reach sizes of billions of elements, rendering conventional static sorting methods insufficient and slow \citep{Batcher1968,Brent1973}.}

\raj{Traditional approaches to parallel sorting, including parallel mergesort and multipass radix sorting, have been extensively studied and optimized over the decades \citep{Batcher1968,Brent1973}. Despite their demonstrated efficacy, the performance of these methods is extremely sensitive to several operational parameters, such as chunk sizes, sorting thresholds, and fallback conditions. These parameters must be tailored to both the specific characteristics of the dataset and the underlying hardware architecture \citep{Shun2015}. Manual tuning of these parameters is laborious and error-prone, and even slight misconfigurations can lead to substantial performance degradation \citep{Brent1973}.}

\raj{To overcome these challenges, our current research has focused on auto-tuning techniques that automatically search for the optimal parameter configurations. Auto-tuning frameworks have been successfully employed to optimize computational kernels and large-scale algorithms by algorithmically exploring multidimensional parameter spaces. Among the various approaches, Genetic Algorithms (GAs) stand out for their robustness, ease of integration, and versatility \citep{10.1162/evco.1993.1.1.1}. By mimicking the process of natural evolution—through selection, recombination , and mutation—GAs are particularly well suited to navigating the complex and high-dimensional spaces that define sorting performance \citep{Holland1975,Goldberg1989,Mitchell1996,Deb2002}.}

\raj{In this paper, we present \textbf{EvoSort}, a general-purpose adaptive sorting framework accessible at the Python level that unifies a refined parallel mergesort and a block-based least significant digit (LSD) radix sort under a GA-driven tuning system. EvoSort is implemented as a Python package, making it immediately accessible to Python developers without requiring low-level programming or system-specific optimizations. The framework automatically and dynamically optimizes critical parameters, such as the insertion sort threshold, fallback thresholds for library-based routines, and even the choice between mergesort and radix sort \citep{10.1093/comjnl/35.6.636}. The design of EvoSort is directly informed by the observation that optimal parameter settings are highly dependent on both data characteristics and the computing environment, a challenge that our GA-based methodology effectively addresses. Importantly, all performance comparisons in this work are conducted against Python-accessible sorting methods (NumPy, PyTorch, Pandas), ensuring that the reported speedups are directly relevant to Python users and represent practical improvements achievable within the Python ecosystem.}

\raj{The core of EvoSort's approach lies in its powerful GA component, which continuously explores a multidimensional parameter space to identify configurations that minimize sorting time while maintaining correctness. Each candidate solution is evaluated by timing the sorting process on a representative dataset, and the best-performing configurations gradually evolve over successive generations \citep{Oliphant2007}. This adaptive mechanism allows EvoSort to respond dynamically to varying data sizes, from tens of millions to billions of elements, and diverse data distributions, thereby achieving significant performance improvements.}

\raj{A noteworthy aspect of EvoSort is its hybrid design. When dealing with integer data types (such as \texttt{int32} or \texttt{int64}), EvoSort can leverage a block-based LSD radix sort that uses bitwise operations to efficiently handle signed integers \citep{Knuth1998,Cormen2009}. However, when the dataset is more amenable to traditional comparison-based sorting, EvoSort employs a refined parallel mergesort that integrates an optimized insertion sort for small subarrays. This seamless switching between sorting strategies, guided by the GA-tuned \texttt{merge\_algorithm} parameter, allows EvoSort to capitalize on the strengths of both approaches, achieving speedups that, in our comprehensive empirical evaluations across diverse data distributions and hardware platforms, range from $0.2\times$ to over $225\times$ compared with conventional routines such as NumPy's \texttt{np.sort} \citep{Oliphant2007}.}

\raj{The adaptive nature of EvoSort, enabled by the evolutionary power of Genetic Algorithms, represents a significant advance in the field of high-performance sorting. It not only provides a robust and scalable solution for sorting extremely large datasets but also exemplifies how metaheuristic optimization techniques can be harnessed to enhance algorithmic performance in modern HPC environments. In essence, EvoSort is more than just a sorting algorithm; it is a template for integrating adaptive auto-tuning strategies into complex computational frameworks.}

The remainder of this paper is organized as follows. Section~\ref{sec:related} reviews related work on parallel sorting and auto-tuning methodologies. Section~\ref{sec:theory} outlines the theoretical foundations of mergesort, LSD radix sorting, and GA-based optimization. Section~\ref{sec:algorithm} details the EvoSort workflow and provides a pseudocode to illustrate its implementation. Section~\ref{sec:setup} describes our experimental setup, including dataset generation and correctness validation. Section~\ref{sec:results} presents comprehensive empirical results for datasets ranging from $10^7$ to $10^{10}$ elements, further Section~\ref{sec:symbolic} takes a step forward and introduces symbolic analysis of the parameters obtained  and Section~\ref{sec:conclusion} concludes with a discussion of our findings and directions for future research.

\section{Related Work}\label{sec:related}

\subsection{Parallel Sorting Approaches}

\raj{Parallel sorting has long been a central research topic in computer science, with early work laying the foundation for efficient algorithm design. Classical approaches, such as the well-known mergesort, divide the dataset recursively and merge sorted subarrays concurrently \citep{Knuth1998,Shun2015}. This divide-and-conquer paradigm has been extended to modern multi-core and distributed architectures, where the ability to perform multiple merge operations simultaneously is crucial for scaling performance.}

\raj{For sorting integer data, radix sort algorithms—both least significant digit (LSD) and most significant digit (MSD) variants—have been extensively explored owing to their linear time complexity under fixed-width assumptions. These algorithms typically rely on histogram-based partitioning techniques that efficiently group elements according to digit values, thereby enabling parallel processing. The inherent simplicity of counting-based partitioning makes these methods particularly attractive for integer sorting, although they require careful management of memory bandwidth and cache usage.}

\raj{Hybrid approaches have also emerged that combine the strengths of different algorithms to minimize overhead. For example, integrating the insertion sort for small subarrays within a larger mergesort framework can significantly reduce the cost of recursive calls and merging operations \citep{Cormen2009}. Such hybrid methods are often effective in practice because they leverage the fast execution of simple algorithms on small datasets, while still benefiting from the scalability of more complex methods on larger datasets.}

\raj{Despite these advancements, many state-of-the-art parallel sorting frameworks still depend on static hand-tuned heuristics. These fixed parameters may work well on specific datasets or hardware platforms but often lack the flexibility to adapt to new environments or varying data characteristics \citep{Cormen2009,Batcher1968}. In high-performance computing (HPC) scenarios, where synchronization costs and memory bandwidth limitations are critical factors, there is a persistent need for sorting algorithms that can dynamically adjust their behavior to minimize overhead and maximize throughput.}

\subsection{Auto-Tuning with Genetic Algorithms}

\raj{Auto-tuning has emerged as a powerful methodology for addressing the limitations of static parameter settings in complex algorithms. These techniques involve systematic exploration of large multidimensional parameter spaces to discover configurations that yield optimal performance. Among various auto-tuning strategies, Genetic Algorithms (GAs) have been particularly successful owing to their robust global search capabilities and their ability to navigate noisy or rugged fitness landscapes \citep{Holland1975,Goldberg1989,Mitchell1996}.}

\raj{GAs operate by mimicking the natural evolutionary process using mechanisms such as selection, recombination , and mutation to iteratively improve a population of candidate solutions. This approach has been applied to a wide range of HPC tasks, including tuning of compiler flags, optimization of memory hierarchies, and determination of block sizes for parallel computations \citep{Mitchell1996,Eiben2003}. Despite these successes, the application of GAs in the domain of parallel sorting remains relatively underexplored.}

\raj{This auto-tuning mechanism not only reduces the burden of manual configuration but also ensures that the sorting process is finely tuned to the specific characteristics of both the dataset and the underlying hardware. The success of this approach highlights the potential of using metaheuristic optimization techniques to drive performance improvements in complex computational systems, setting a new benchmark for adaptive sorting in HPC environments.}

\vspace{1ex}
\noindent \raj{In summary, while classical parallel sorting algorithms have laid a robust foundation for high-performance data processing, the integration of auto-tuning strategies—especially those based on Genetic Algorithms—represents a significant step forward in addressing the challenges of modern, large-scale sorting. EvoSort leverages these advances to provide a flexible, high-performance sorting framework that dynamically optimizes its behavior, thereby ensuring robust performance across a wide range of applications and computing platforms.}

\section{Theoretical Foundations}\label{sec:theory}

\subsection{Parallel Mergesort and Radix Sort Complexity}

\paragraph*{Parallel Mergesort:}
A classical mergesort is renowned for its $O(n \log n)$ time complexity, which is achieved through a recursive divide-and-conquer strategy that splits the array into two halves, sorts each half independently, and then merges the sorted subarrays into a single sorted sequence \citep{Knuth1998,Cormen2009}. In a parallel computing environment, a mergesort can leverage concurrent processing by distributing the merge operations across multiple processors. However, the theoretical benefits of parallelization are often tempered by practical challenges, such as synchronization overhead, load imbalance, and memory bandwidth limitations \citep{Shun2015,Batcher1968}. 

In many high-performance applications, it has been observed that as subarrays become smaller, the overhead of recursive calls and parallel task management can outweigh the benefits of parallel execution. Consequently, many implementations adopt a hybrid approach by switching to simpler algorithms, such as insertion sort, for small subarrays. This hybridization enhances cache performance and reduces constant factors in execution time \citep{Cormen2009,Knuth1998}. Such considerations are vital in modern HPC environments, where the efficient use of resources is crucial for handling large-scale datasets.

\paragraph*{Block-Based LSD Radix Sort:}
Radix Sort is particularly effective for sorting fixed-width integer data, where its linear time complexity becomes a significant advantage. The least significant digit (LSD) variant of radix sort processes numbers digit by digit, typically requiring only a few passes (e.g., four passes for 32-bit integers when processing 8 bits per pass) \citep{Brent1973,Sanders2004}. This method uses the histogram-based partitioning of group numbers based on individual digit values, thereby enabling parallel processing. 

Handling negative numbers in radix sorting requires additional care. A common technique is to perform bitwise transformations that effectively map the signed range into the unsigned domain, preserving the natural order of numbers \citep{Knuth1998,Cormen2009}. Moreover, in high-performance settings, radix sort implementations often utilize thread-local histograms to reduce contention, ensuring that each thread can operate independently before combining the results. This approach minimizes synchronization overhead and maximizes throughput in multicore architectures \citep{Sanders2004,Shun2015}.

\subsection{GA Convergence and Optimization}

Genetic Algorithms (GAs) have emerged as a leading approach for solving complex optimization problems, especially in cases where the search space is high-dimensional, non-linear, and discontinuous. A GA evolves a population of candidate solutions over successive generations using biologically inspired operators such as selection, recombination, and mutation \citep{Holland1975,Goldberg1989,Mitchell1996}. 

In the context of EvoSort, each candidate solution is represented as a vector \raj{of key parameters of the proposed sorting procedure:}
\[
\mathbf{x} \;=\; (\,T_{\text{insertion}},\, A_{\text{code}},\, T_{\text{numpy}}),
\]
where:
\begin{itemize}
  \item $T_{\text{insertion}}$ is the threshold for switching to insertion sort,
  \item $A_{\text{code}}$ is a discrete parameter that selects between different algorithmic strategies (e.g., choosing between mergesort and LSD radix sort), and
  \item $T_{\text{numpy}}$ defines the threshold at which NumPy's built-in sort is invoked.
\end{itemize}

The fitness of each candidate was measured using the sorting time on a representative dataset, and the goal was to minimize this time \citep{Oliphant2007}. Through iterative selection, recombination , and mutation, the GA refines the population of candidate configurations, gradually converging to near-optimal parameter settings that balance the trade-offs between parallel efficiency and algorithmic overhead \citep{Holland1975,Goldberg1989,Mitchell1996,Eiben2003,Baeck1996}.

Our work builds on the foundational contributions to evolutionary multi-objective optimization, particularly in the context of genetic algorithms, as demonstrated in earlier research \citep{Deb2001,Deb2002}. These studies illustrate how evolutionary strategies can be effectively adapted to address complex optimization problems involving multiple conflicting objectives. Although EvoSort is primarily focused on optimizing a single performance metric (sorting time), the principles derived from multi-objective optimization research, such as maintaining diversity in the population and avoiding premature convergence, are crucial in ensuring that the GA explores a rich variety of parameter settings. This influence is evident in EvoSort’s design, where the GA robustly searches for the parameter space, balancing exploration and exploitation to dynamically adapt to changing data characteristics and hardware environments \citep{Deb2001,Deb2002}.

In summary, the theoretical foundation of EvoSort is based on two pillars. First, the complexity analysis of parallel mergesort and block-based LSD radix sorting established the performance characteristics and potential bottlenecks of these algorithms. Second, the robust convergence properties of Genetic Algorithms, as established in prior work \citep{Deb2001,Deb2002}, offer a powerful mechanism for automatically tuning sorting parameters to achieve optimal performance in high-performance computing contexts.

\section{EvoSort Algorithm}\label{sec:algorithm}

\raj{EvoSort is a general-purpose adaptive sorting framework implemented as a Python package that combines the efficiency of parallel sorting techniques with the robust optimization capabilities of Genetic Algorithms (GAs). As a Python-accessible framework, EvoSort provides a simple, high-level interface that Python developers can use directly, similar to NumPy's \texttt{np.sort()} or other standard Python sorting routines. Unlike traditional sorting methods that rely on fixed or manually tuned parameters, EvoSort dynamically adapts to both the size of the data and the hardware characteristics. By automatically tuning key parameters through evolutionary computation, EvoSort selects the most effective strategy—whether it is a refined parallel mergesort or block-based LSD radix sort—based on the outcome of its GA-driven search. This approach builds on classical algorithmic research \citep{Knuth1998,Cormen2009,Batcher1968} and modern advances in evolutionary computations \citep{Holland1975,Goldberg1989,Mitchell1996,Deb2001,Deb2002}.}

In the following subsections, we describe the major components of EvoSort in detail, explain the underlying mathematics, and the pseudocodes.

\subsection{EvoSort Master Pipeline}

The master pipeline orchestrates the entire process by managing sorting tasks across a range of dataset sizes, from tens of millions to tens of billions of elements. For each dataset size, the pipeline performs the following steps.
\begin{enumerate}
  \item \textbf{GA-Driven Tuning:} The GA is invoked to optimize the parameter set that governs the sorting behavior. This tuning process finds near-optimal values for  
    \begin{itemize}
      \item \emph{Insertion Sort Threshold} ($T_{\text{insertion}}$),
      \item \emph{Merge Algorithm Code} ($A_{\text{code}}$), and
      \item \emph{NumPy Sort Threshold} ($T_{\text{numpy}}$).
    \end{itemize}
  \item \textbf{Data Generation:} A data array $A$ is created with random integers, typically drawn from a uniform distribution over \([-10^9, +10^9]\).
  \item \textbf{Reference Sorting:} A reference sorted array is computed using NumPy’s \verb|np.sort| to ensure correctness.
  \item \textbf{Final Sorting:} The optimized parameters are applied in the Adaptive Partition Sort subroutine to sort the full dataset.
  \item \textbf{Validation:} The output is validated against the reference sort to ensure correctness.
\end{enumerate}

Algorithm~\ref{alg:master} provides a high-level pseudocode view of the master pipeline.

\begin{algorithm}[ht]
\caption{\textbf{EvoSort\_MasterPipeline}}
\label{alg:master}
\begin{algorithmic}[1]
\REQUIRE \textbf{Sizes} = array sizes of interest, e.g., $\{10^7,\,10^8,\,5\times10^8,\dots\}$
\FOR{\textbf{each} $n$ \textbf{in} Sizes}
  \STATE $bestParams \leftarrow \text{RunGATuning}(n)$ \COMMENT{Optimize parameters for dataset size $n$ \citep{,Deb2001}}
  \STATE Create data array $A$ of size $n$ with random integers 
  \STATE $\text{RefSorted} \leftarrow \text{np.sort}(A)$ \COMMENT{Reference sort for correctness \citep{Oliphant2007}}
  \STATE $A_{\text{Evo}} \leftarrow \text{AdaptivePartitionSort}(A, bestParams)$
  \STATE \textbf{assert}($A_{\text{Evo}}$ equals RefSorted) \COMMENT{Ensures sorting correctness}
  \STATE Compare runtime vs.\ \texttt{np.sort}(kind='quicksort'), \texttt{np.sort}(kind='mergesort'), etc.
\ENDFOR
\end{algorithmic}
\end{algorithm}

\subsection{RunGATuning: GA Driver}

\raj{At the heart of EvoSort is the GA driver, which automatically optimizes three key parameters. Each candidate solution is represented as a vector:
\[
\mathbf{x} = \Bigl( T_{\text{insertion}},\; A_{\text{code}},\; T_{\text{numpy}} \Bigr),
\]
where each component controls a critical aspect of the sorting process. The GA starts with a population of 30 individuals and iteratively improves the solutions through standard operations (selection, recombination, and mutation) over 10 generations. The parameter bounds are defined as follows: $T_{\text{insertion}} \in [2, 2048]$ (with initial generation bounds $[2, 256]$), $A_{\text{code}} \in \{3, 4\}$ where 3 represents mergesort and 4 represents LSD radix sort, and $T_{\text{numpy}} \in [1000, 100000]$ (with initial generation bounds $[30000, 100000]$). The fitness of each candidate is measured by the wall-clock sorting time on a sample dataset. Mathematically, the fitness function is defined as
\[
f(\mathbf{x}) = T_{\text{sort}}(\mathbf{x}),
\]
and the GA seeks to minimize \( f(\mathbf{x}) \). The GA uses uniform crossover with probability 0.7 and uniform mutation with probability 0.3, with elitism preserving the best individuals across generations. The convergence of the GA is typically achieved within 10 generations, as evidenced by progressively tighter convergence in the average execution time across generations \citep{Holland1975,Goldberg1989,Mitchell1996,Deb2002}.}

Algorithm~\ref{alg:gaTuning} summarizes the GA driver routine.

\begin{algorithm}[h]
\caption{\textbf{RunGATuning}($n$)}
\label{alg:gaTuning}
\begin{algorithmic}[1]
\STATE \textbf{Input:} $n$, the dataset size to tune
\STATE \text{sampleData} $\leftarrow$ random integer array of size $n$ (uniform distribution, \texttt{int32})
\STATE Initialize population $P$ with 30 individuals, each with 3 parameters: $(T_{\text{insertion}}, A_{\text{code}}, T_{\text{numpy}})$
\FOR{$g=0$ to $9$}
   \STATE Evaluate each candidate's fitness via timed sorting on \texttt{sampleData}
   \STATE Select best individuals (elitism)
   \STATE Apply uniform crossover with probability 0.7
   \STATE Apply uniform mutation with probability 0.3
   \STATE Re-evaluate invalidated individuals
\ENDFOR
\STATE \textbf{Return} best individual's parameter set
\end{algorithmic}
\end{algorithm}

Next, we wanted to highlight the refined parallel merge sort, as evident in Algorithm~\ref{alg:refinedMergeSort}, and how it is different from a standard implementation. Traditional merge sort performs recursive division followed by merging, often leading to significant overhead due to deep recursion and non-parallel stages. In contrast, our version adopts a bottom-up strategy that begins with in-place insertion sort on small base chunks to exploit cache locality. These sorted chunks are then progressively merged in a parallel fashion, using fixed-size buffers and batch-wise coordination across threads. This not only minimizes memory allocations but also enables better thread scalability by reducing contention during merging. The combination of localized sorting and staged parallel merges results in a significantly more efficient and high-throughput variant tailored for modern multi-core architectures.

\begin{algorithm}[H]
\caption{\textbf{RefinedParallelMergeSort}($A$, $chunkSize$)}
\label{alg:refinedMergeSort}
\begin{algorithmic}[1]
\STATE \textbf{Input:} $A$, input array of size $n$; $chunkSize$, the base size for insertion sort
\STATE Partition $A$ into chunks of size $chunkSize$
\FORALL{chunk $C$ in parallel}
    \STATE Perform \texttt{InsertionSort} on $C$
\ENDFOR
\STATE $currSize \leftarrow chunkSize$
\WHILE{$currSize < n$}
    \STATE $step \leftarrow 2 \times currSize$
    \FORALL{pairs of chunks of size $currSize$ in parallel}
        \STATE Perform optimized merge using \texttt{MergeStandardOpt}
    \ENDFOR
    \STATE $currSize \leftarrow 2 \times currSize$
\ENDWHILE
\STATE \textbf{Return} sorted array $A$
\end{algorithmic}
\end{algorithm}

Further, EvoSort also includes a highly optimized block-based Least Significant Digit (LSD) radix sort, designed specifically for signed integer types. As shown in Algorithm~\ref{alg:refinedRadixSort} and Algorithm~\ref{alg:refinedRadixSort64}, this approach begins by applying a bitwise XOR transformation to the input data, effectively converting signed integers into an unsigned space while preserving correct ordering. The algorithm then performs multiple passes—one byte at a time—building local histograms per thread in parallel. These histograms are combined into global prefix sums to determine final write positions. Each thread independently redistributes its chunk of data into a temporary buffer based on byte-level radix keys, and input/output buffers are swapped after every pass. This strategy enables efficient load balancing, avoids branching, and leverages high memory bandwidth to deliver fast, scalable performance on large datasets.

While both the 32-bit and 64-bit implementations follow the same high-level structure, they differ in the number of passes and the bit manipulations required. The 32-bit version performs four radix passes—each handling 8 bits—sufficient to cover all bits of an \texttt{int32}. In contrast, the 64-bit variant performs eight passes to fully process the wider 64-bit integer space. The XOR transformation at the start of each function also uses different masks (\texttt{0x80000000} for 32-bit vs. \texttt{0x8000000000000000} for 64-bit) to properly flip the sign bit for correct signed-to-unsigned mapping. Due to the increased number of passes and larger data size, the 64-bit version requires more memory bandwidth and compute per element but retains the same parallel and chunked structure for efficiency.

\begin{algorithm}[H]
\caption{\textbf{RefinedParallelRadixSort\_int32}($A$)}
\label{alg:refinedRadixSort}
\begin{algorithmic}[1]
\STATE \textbf{Input:} $A$, array of signed 32-bit integers
\STATE \texttt{XOR} each element with \texttt{0x80000000} to handle negatives
\FOR{$pass=0$ to $3$}
    \STATE $shift \leftarrow 8 \times pass$
    \STATE Compute local histograms for each thread in parallel
    \STATE Reduce to global histogram and compute prefix sums
    \STATE Compute thread-wise write offsets
    \FORALL{elements in parallel}
        \STATE Use offset to place element in temporary array
    \ENDFOR
    \STATE Swap input and temporary array
\ENDFOR
\STATE \texttt{XOR} each element again with \texttt{0x80000000} to restore original values
\STATE \textbf{Return} sorted array $A$
\end{algorithmic}
\end{algorithm}

\begin{algorithm}[H]
\caption{\textbf{RefinedParallelRadixSort\_int64}($A$)}
\label{alg:refinedRadixSort64}
\begin{algorithmic}[1]
\STATE \textbf{Input:} $A$, array of signed 64-bit integers
\STATE \texttt{XOR} each element with \texttt{0x8000000000000000} to handle negatives
\FOR{$pass=0$ to $7$}
    \STATE $shift \leftarrow 8 \times pass$
    \STATE Compute local histograms for each thread in parallel
    \STATE Reduce to global histogram and compute prefix sums
    \STATE Compute thread-wise write offsets
    \FORALL{elements in parallel}
        \STATE Use offset to place element in temporary array
    \ENDFOR
    \STATE Swap input and temporary array
\ENDFOR
\STATE \texttt{XOR} each element again with \texttt{0x8000000000000000} to restore original values
\STATE \textbf{Return} sorted array $A$
\end{algorithmic}
\end{algorithm}

\subsection{Adaptive Partition Sort Subroutine}

Once the optimal parameters are determined, EvoSort applies the Adaptive Partition Sort subroutine to sort the dataset. This subroutine makes an intelligent decision based on current parameters.
\begin{itemize}
  \item If the array size is below $T_{\text{numpy}}$, it falls back on NumPy’s highly optimized sort.
  \item Otherwise, if the algorithm code \(A_{\text{code}} = 4\) and the data type is integer, it invokes the block-based LSD radix sort as detailed in \ref{alg:refinedRadixSort} and \ref{alg:refinedRadixSort64} for the respective types.
  \item For other cases, it uses a refined parallel mergesort that integrates insertion sort for small subarrays.
\end{itemize}
This dynamic selection allows EvoSort to optimize its performance across varying data sizes and distributions \citep{Brent1973,Sanders2004}.

Algorithm~\ref{alg:APS} details this subroutine.

\begin{algorithm}[h]
\caption{\textbf{AdaptivePartitionSort}}
\label{alg:APS}
\begin{algorithmic}[1]
\REQUIRE Array $A$; parameters: $T_{\text{insertion}}$, $A_{\text{code}}$, $T_{\text{numpy}}$
\IF{$|A| < T_{\text{numpy}}$}
   \STATE $A.\text{sort}()$
\ELSIF{$A_{\text{code}} = 4$ \textbf{and} array is integer-typed}
   \STATE \texttt{ParallelLSDRadixSort}($A$) \COMMENT{Effective for large integer arrays \citep{Brent1973,Sanders2004}}
\ELSIF{$A_{\text{code}} = 3$}
   \STATE \texttt{RefinedParallelMergesort}($A, T_{\text{insertion}}$)
\ELSE
   \STATE \texttt{RefinedParallelMergesort}($A, T_{\text{insertion}}$)
\ENDIF
\RETURN $A$
\end{algorithmic}
\end{algorithm}

Each branch is designed to minimize overhead by dynamically switching between algorithms based on the input size and data type.

\subsection{Overall Workflow Diagram}

\begin{figure}[ht]
  \centering
  \includegraphics[width=0.5\textwidth]{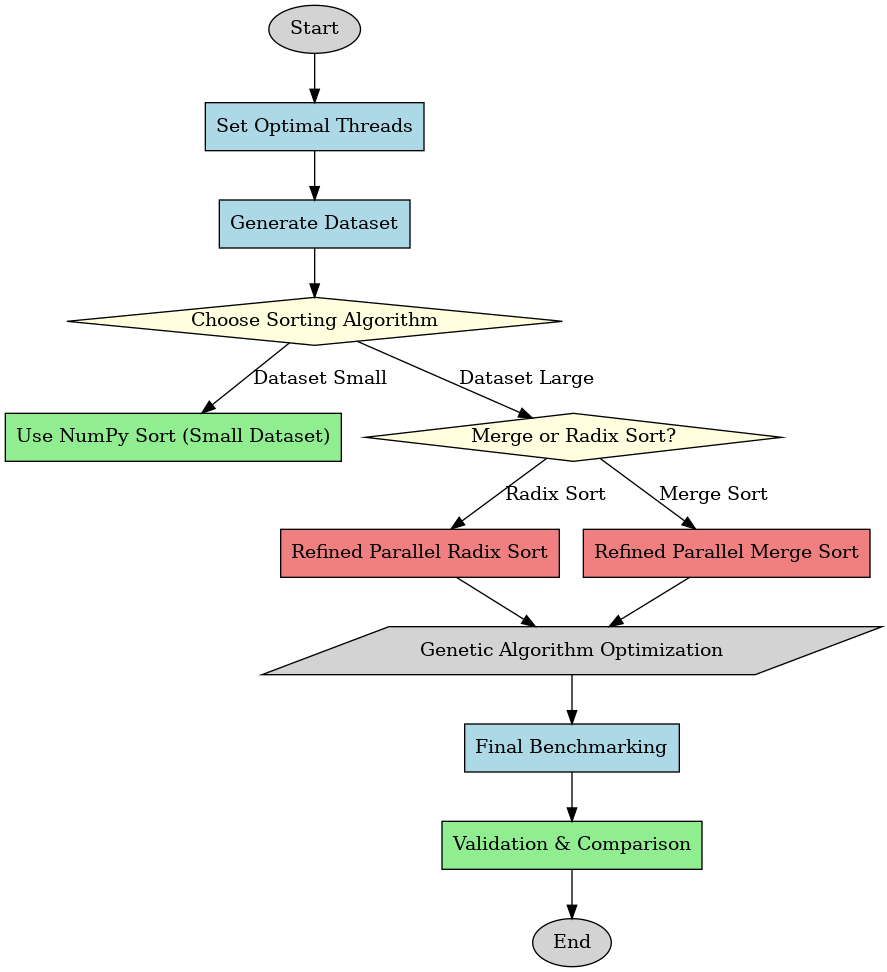}
  \caption{EvoSort Workflow. The process begins by setting the optimal threading and generating a dataset. If the dataset is small, EvoSort directly uses NumPy's built-in sort. Otherwise, it selects between a refined parallel mergesort or a block-based LSD radix sort (when \texttt{merge\_algorithm == 4}), both of which are tuned by a Genetic Algorithm. After the final sorting is completed, the results are validated against a reference sort (typically \texttt{np.sort})}
  \label{fig:evo_workflow}
\end{figure}

\noindent
\raj{To provide a clear overview of EvoSort's entire process, Figure~\ref{fig:evo_workflow} shows the data flow from the initial setup (optimal threading and dataset generation) to the final sorting stage. Specifically, if the dataset is small, EvoSort relies on NumPy's highly optimized sort. Otherwise, it determines whether to use a refined parallel mergesort or a block-based LSD radix sort based on parameters tuned by a Genetic Algorithm (GA). The GA itself iterates over possible parameter values, measuring sorting time to guide its search. Once sorting is complete, EvoSort compares the output against a reference array, ensuring correctness. This adaptive mechanism enables EvoSort to consistently achieve high performance across diverse data sizes and types.}

\subsection{Summary}

\raj{EvoSort's architecture and pseudocode have been carefully designed to integrate a robust GA-driven parameter tuning mechanism with efficient parallel sorting algorithms. The GA optimizes crucial parameters by minimizing the sorting time, as expressed by
\[
\min_{\mathbf{x}} f(\mathbf{x}) = T_{\text{sort}}(\mathbf{x}),
\]
and converges in a small number of generations. Our detailed pseudocode and the comprehensive workflow diagram in Figure~\ref{fig:evo_workflow} illustrate the seamless integration of data generation, parameter tuning, algorithm selection, and final output validation. This self-tuning strategy sets EvoSort apart from conventional static methods and demonstrates significant novelty and scalability for modern high-performance computing applications \citep{Holland1975,Goldberg1989,Mitchell1996,Deb2001,Deb2002,Oliphant2007}.}


\section{Experimental Setup}\label{sec:setup}

\raj{Our evaluation of EvoSort is designed to mimic real-world high-performance computing (HPC) usage while providing rigorous, quantitative comparisons. It is important to note that EvoSort is positioned as a general-purpose sorting framework accessible at the Python level, and all benchmarks in this work compare EvoSort against other Python-accessible sorting methods. This ensures that the reported performance improvements are directly applicable to Python users and represent practical speedups achievable within the Python ecosystem, rather than comparisons against low-level C/C++ implementations that are not directly accessible to Python developers.} In this section, we describe the hardware environment, dataset generation methodology, and baseline comparisons in detail, along with the mathematical framework for performance evaluation.

\paragraph*{Hardware Environment:}  
The experiments are executed on a high-end HPC node equipped with 1024\,GB of RAM and a multi-core processor configuration. Such extensive memory capacity allows us to hold arrays of up to $10^{10}$ elements entirely in main memory, eliminating the variability introduced by disk I/O. All sorting routines are implemented in Python 3.6.8, and performance-critical sections are accelerated using Numba's just-in-time (JIT) compilation \citep{Lam2015}. This acceleration ensures that the Python overhead is minimized and that the algorithms are executed close to native speed. Moreover, our approach leverages implicit parallelization within the refined parallel mergesort and block-based LSD radix sort algorithms, without requiring explicit concurrency flags, thereby simplifying deployment while maximizing resource utilization \citep{Frigo1999}.

\paragraph*{Dataset Generation:}  
\raj{To simulate realistic HPC workloads, datasets are generated using \texttt{np.random.randint} with \texttt{dtype=np.int32}, creating 32-bit signed integer arrays with values uniformly distributed in the interval $[-10^9, +10^9]$. All experimental data throughout this work uses \texttt{int32} (4 bytes per element), ensuring consistent memory usage and enabling fair comparisons across all algorithms. The sizes of these arrays range from $10^7$ to $10^{10}$ elements.}

\paragraph*{Baseline Comparisons:}  
\raj{To evaluate the performance of EvoSort as a general-purpose Python-accessible sorting framework, we benchmark it against established sorting routines that are directly accessible to Python developers. All comparisons are conducted at the Python level, ensuring that the reported speedups represent practical improvements achievable within the Python ecosystem. Specifically, we compare EvoSort against:}
\begin{itemize}
  \item \raj{\texttt{np.sort} with \texttt{kind='quicksort'} (the default sorting algorithm in NumPy), representing the most commonly used Python sorting method in scientific computing,}
  \item \raj{\texttt{np.sort} with \texttt{kind='mergesort'}, NumPy's stable sorting algorithm,}
  \item \raj{\texttt{np.sort} (default), NumPy's default sorting method without specifying the algorithm,}
  \item \raj{PyTorch CPU sort (\texttt{torch.sort}), representing state-of-the-art sorting performance in the deep learning ecosystem, and}
  \item \raj{Pandas sort (\texttt{pd.Series.sort\_values}), representing common data science workflows.}
\end{itemize}
\raj{This comprehensive comparison against Python-accessible methods ensures that all reported speedups are directly relevant to Python users and represent practical improvements that can be immediately realized by replacing existing Python sorting calls with EvoSort.} The primary performance metric is the wall-clock runtime, measured in seconds. We compute the speedup factor $S$ of EvoSort relative to a baseline as:
\[
S = \frac{T_{\text{baseline}}}{T_{\text{EvoSort}}},
\]
where $T_{\text{baseline}}$ is the time taken by the Python-accessible baseline and $T_{\text{EvoSort}}$ is the time taken by EvoSort. These mathematical comparisons provide an objective basis for evaluating the performance improvements achieved by EvoSort \citep{,Sutter2005}.

\raj{In summary, our experimental setup is carefully designed to reflect realistic HPC usage while maintaining focus on Python-accessible comparisons. By comparing EvoSort against well-established Python-accessible sorting routines and using mathematical formulations to quantify improvements, we provide a comprehensive and accurate assessment of EvoSort's capabilities that is directly relevant to Python developers.}

\section{Results and Analysis}\label{sec:results}

\raj{This section presents a comprehensive analysis of EvoSort's performance across datasets ranging from $10^7$ to $10^{10}$ elements. We first provide a unified overview of the GA convergence process, then present detailed results for each dataset size with corresponding figures, followed by comprehensive tables summarizing all performance metrics and optimized parameters.}

\subsection{GA Convergence Process Overview}

\raj{For each dataset size, the GA optimization process follows a consistent pattern. The GA starts with a population of 30 individuals that explores various combinations of the three critical parameters: insertion sort threshold ($T_{\text{insertion}}$), merge algorithm code ($A_{\text{code}}$), and NumPy sort threshold ($T_{\text{numpy}}$). The parameter bounds are: $T_{\text{insertion}} \in [2, 256]$ for Generation~0 (expanded to $[2, 2048]$ for subsequent generations), $A_{\text{code}} \in \{3, 4\}$ where 3 represents mergesort and 4 represents LSD radix sort, and $T_{\text{numpy}} \in [30000, 100000]$ for Generation~0 (expanded to $[1000, 100000]$ for subsequent generations). Over 10 generations, the GA applies uniform crossover (with probability 0.7) and uniform mutation (with probability 0.3) while preserving the best individuals via elitism \citep{Holland1975,Goldberg1989,Mitchell1996,Deb2001,Deb2002}. The fitness of each candidate is measured by the wall-clock sorting time on a representative dataset, and the GA seeks to minimize this time.}

\raj{The GA convergence process typically exhibits three key characteristics across all dataset sizes: (1) \textbf{Initial Population Diversity}, where Generation~0 shows a wide range of execution times (with worst-to-best ratios ranging from 21$\times$ to 98$\times$ across different dataset sizes) reflecting the sensitivity of sorting performance to parameter choices; (2) \textbf{Rapid Improvement}, where the GA quickly refines the population through selection, recombination, and mutation, with most configurations converging to near-optimal ranges within 2--6 generations; and (3) \textbf{Stable Convergence}, where the best solutions remain consistent in later generations (typically by Generation~5--9), indicating that near-optimal parameter sets have been identified. The following subsections present detailed convergence results and optimized parameters for each dataset size.}

\subsection{Performance Comparison Across Data Sizes}

\subsection{Genetic Algorithm Convergence at 10 Million Elements}

\begin{figure}[h]
    \centering
    \includegraphics[width=0.5\textwidth]{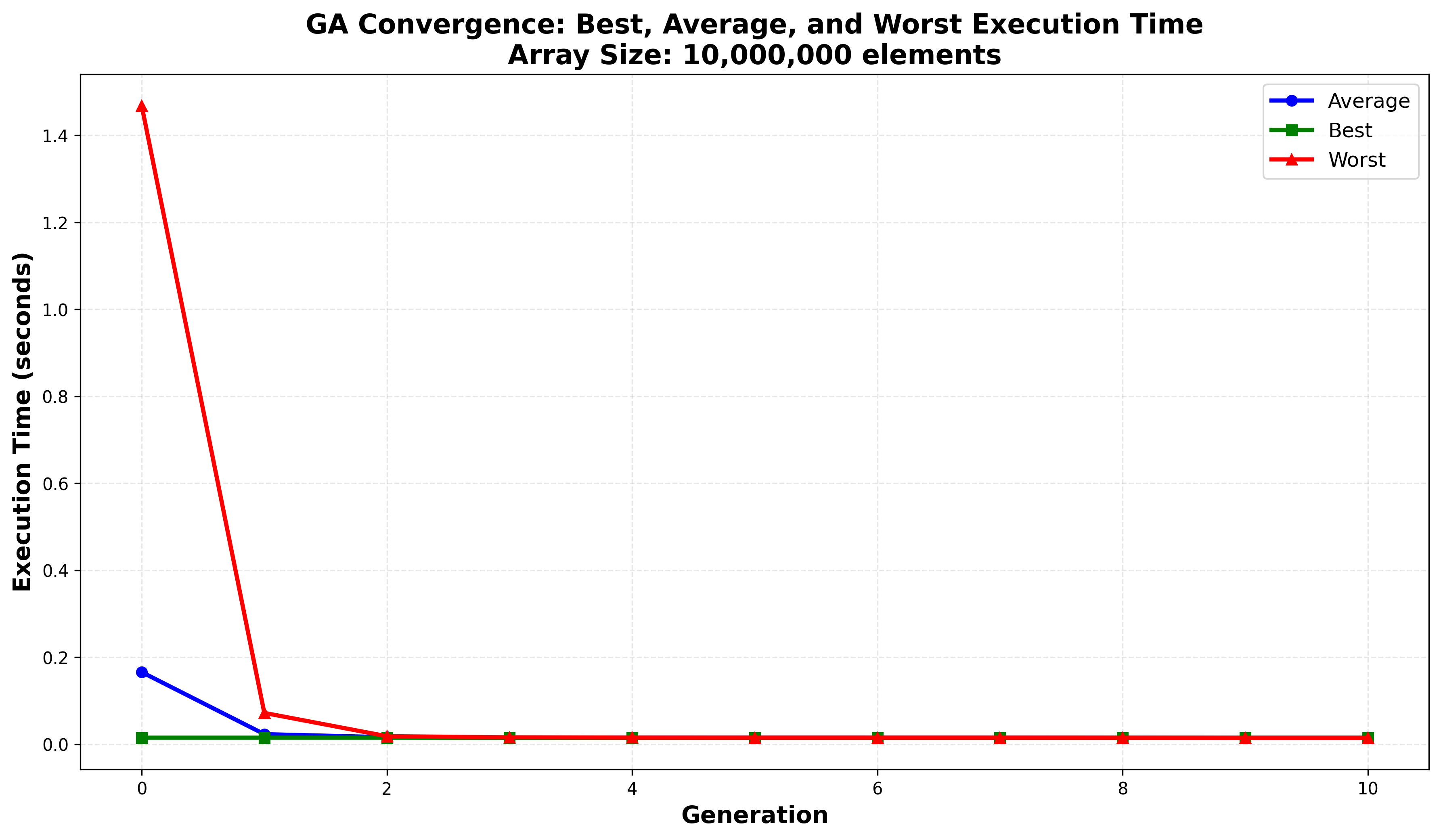}
    \caption{\raj{GA Optimization for a 10\,M-element Array. The plot shows the GA's progression over 10 generations (best, worst, and average execution times). The GA converges to optimal parameters within 3 generations, with the best solution achieving 0.0149\,s.}}
    \label{fig:ga_10M}
\end{figure}

\begin{figure}[h]
    \centering
    \includegraphics[width=0.5\textwidth]{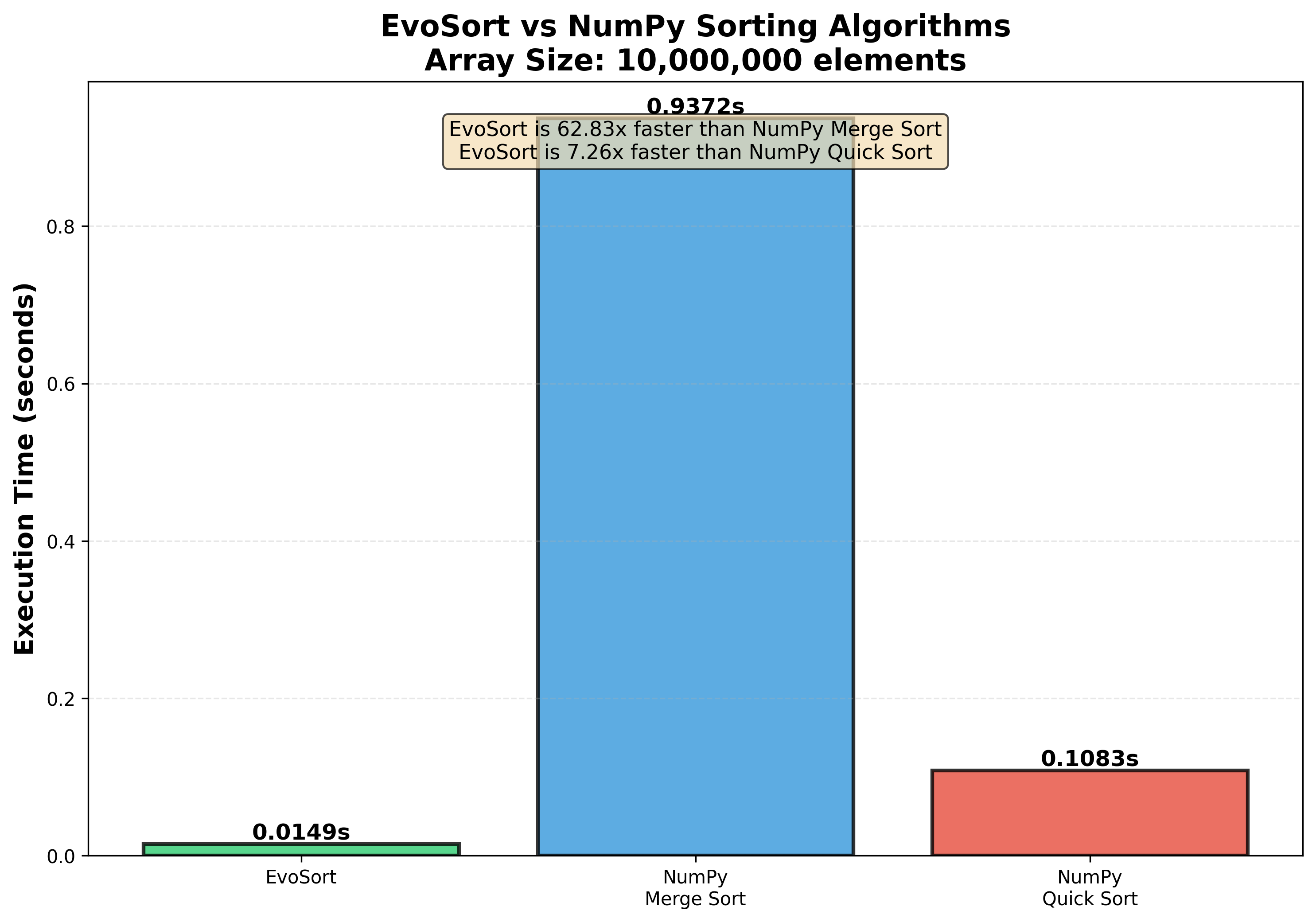}
    \caption{\raj{Performance Comparison for a 10\,M-element Array. EvoSort (0.0149\,s) outperforms NumPy quicksort (0.1083\,s) by 7.3$\times$ and mergesort (0.9372\,s) by 62.8$\times$.}}
    \label{fig:comp_10M}
\end{figure}

\raj{For the 10 million (\(10^7\)) element dataset, the GA quickly reduces the average sorting time from 0.166\,s in Generation~0 to 0.015\,s by Generation~3, with the best solution achieving 0.0149\,s from Generation~3 onwards. The optimized parameters are \([629,\, 4,\, 78447]\), where the insertion sort threshold is 629, the merge algorithm code is 4 (LSD radix sort), and the NumPy sort threshold is 78,447. The final EvoSort runtime of 0.0149\,s outperforms NumPy quicksort (0.1083\,s) by 7.3$\times$ and mergesort (0.9372\,s) by 62.8$\times$.}

\subsection{Genetic Algorithm Convergence at 100 Million Elements}

\begin{figure}[h]
    \centering
    \includegraphics[width=0.5\textwidth]{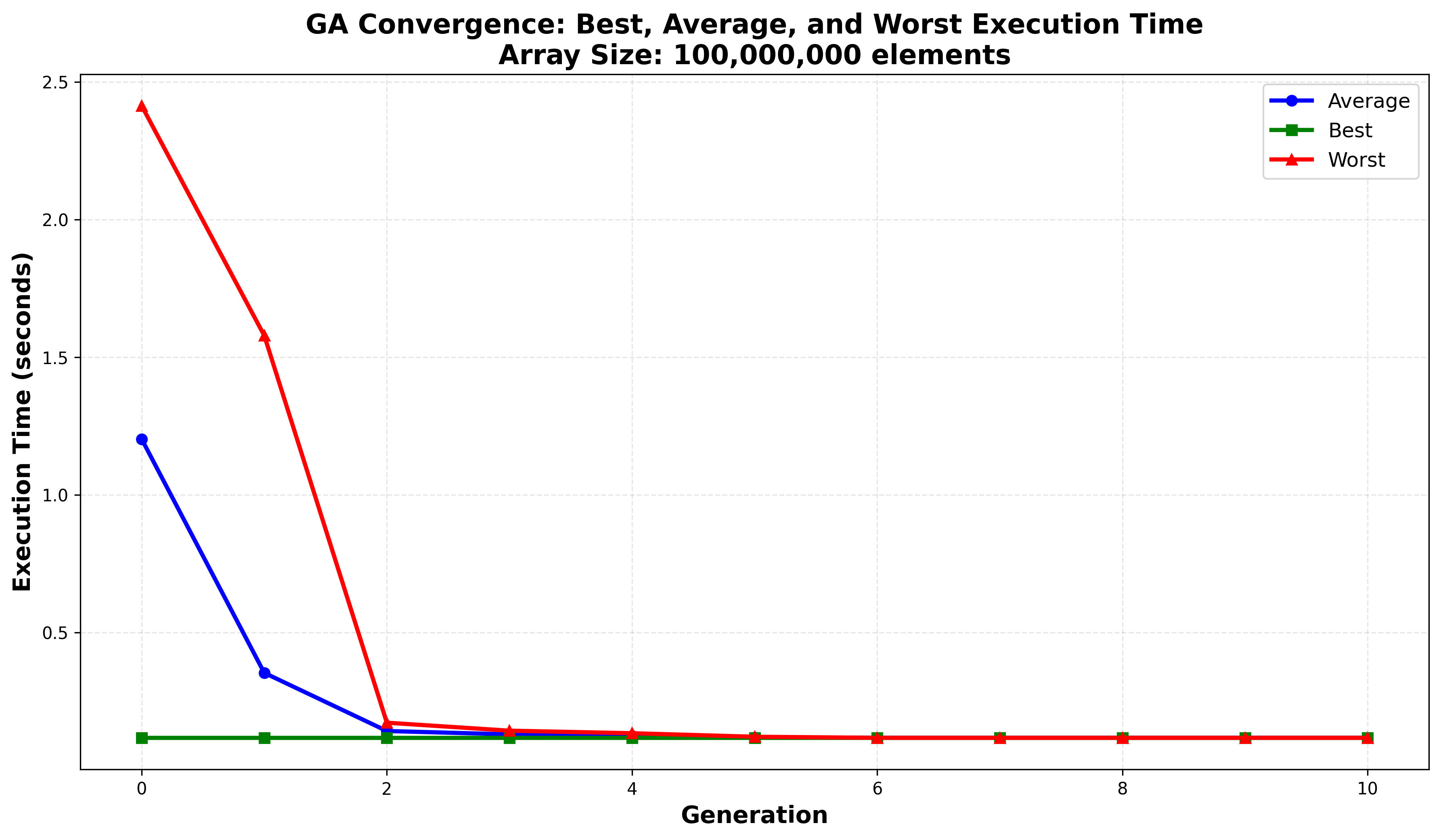}
    \caption{\raj{GA Optimization for a 100\,M-element Array. The plot shows the GA's progression over 10 generations. The GA finds the optimal solution in Generation~0, with the average converging by Generation~5, achieving 0.1174\,s.}}
    \label{fig:ga_100M}
\end{figure}

\begin{figure}[h]
    \centering
    \includegraphics[width=0.5\textwidth]{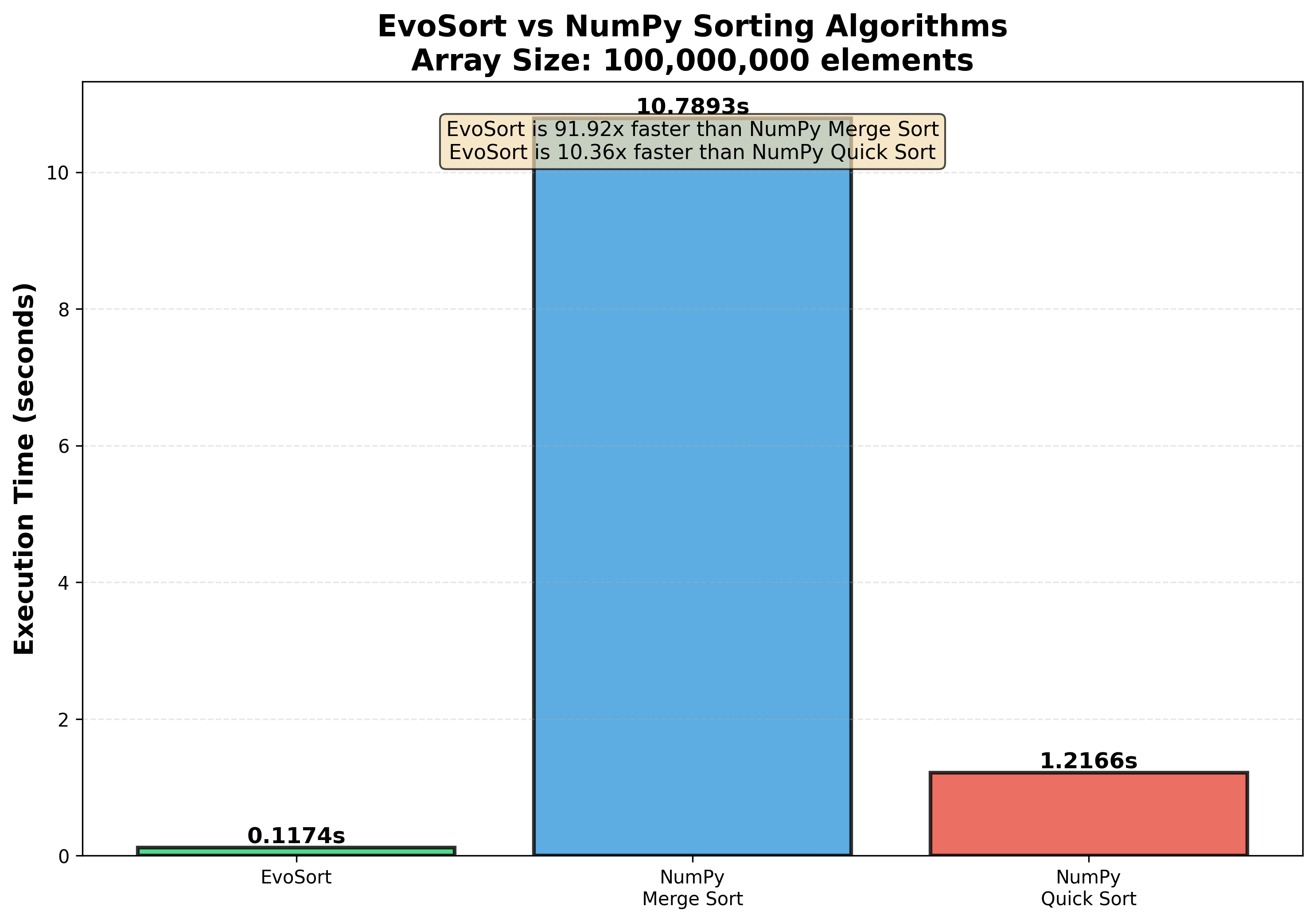}
    \caption{\raj{Performance Comparison for a 100\,M-element Array. EvoSort (0.1174\,s) outperforms NumPy quicksort (1.2166\,s) by 10.4$\times$ and mergesort (10.7893\,s) by 91.9$\times$.}}
    \label{fig:comp_100M}
\end{figure}

\raj{For the 100 million (\(10^8\)) element dataset, the GA converges to optimal parameters \([81,\, 4,\, 32199]\) with execution times ranging from 2.41\,s (worst configuration) down to 0.1174\,s (best candidate) in Generation~0, with the best solution appearing immediately in Generation~0 and the average converging by Generation~5. The optimized parameters are: insertion sort threshold of 81, merge algorithm code of 4 (LSD radix sort), and NumPy sort threshold of 32,199. The final EvoSort runtime of 0.1174\,s outperforms NumPy quicksort (1.2166\,s) by 10.4$\times$ and mergesort (10.7893\,s) by 91.9$\times$.}

\subsection{Genetic Algorithm Convergence at 500 Million Elements}

\begin{figure}[h]
    \centering
    \includegraphics[width=0.5\textwidth]{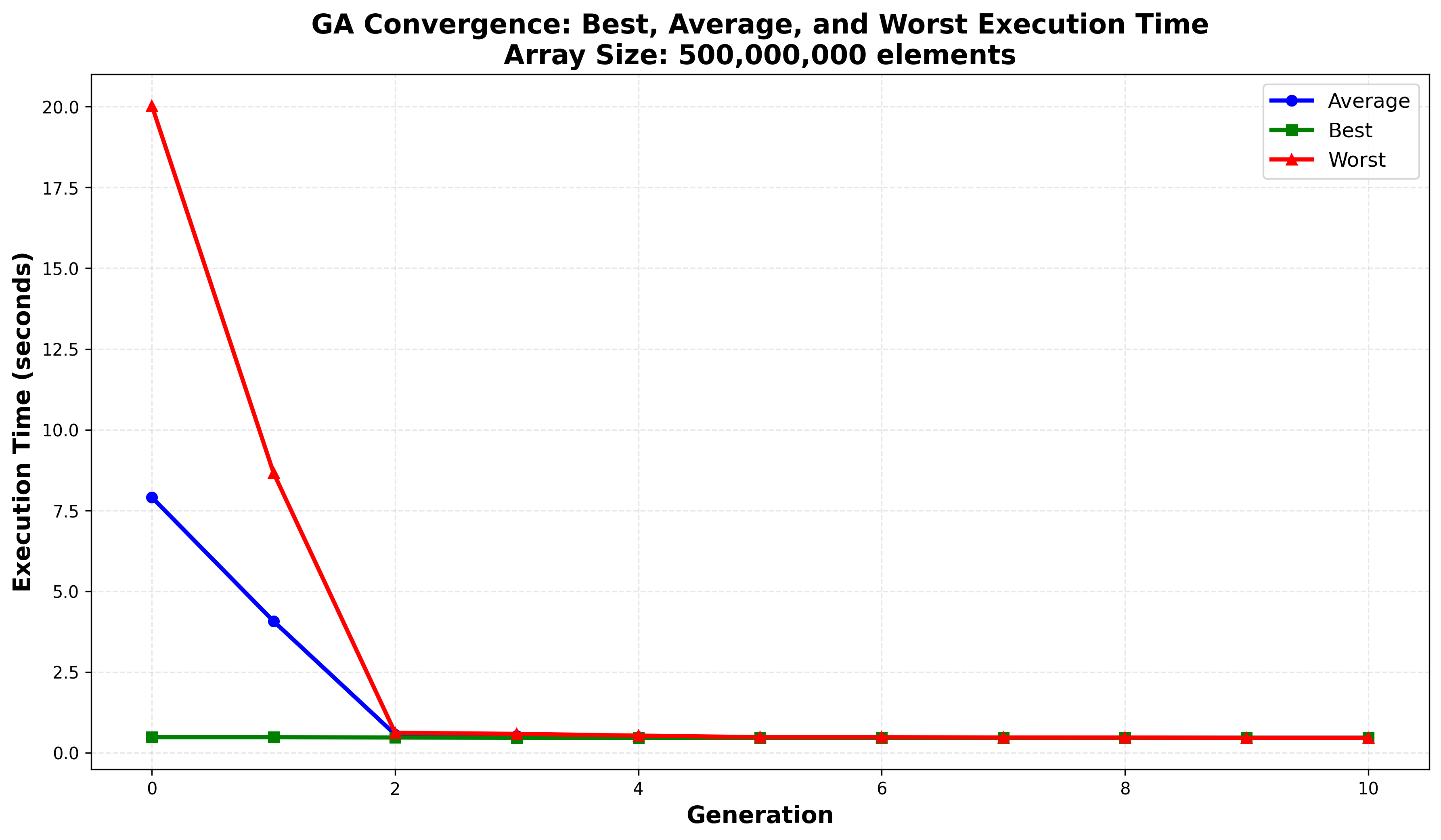}
    \caption{\raj{GA Optimization for a 500\,M-element Array. The plot displays the evolution of the GA over 10 generations. The GA finds the optimal solution by Generation~3, with the average converging by Generation~5, achieving 0.4655\,s.}}
    \label{fig:ga_500M}
\end{figure}

\begin{figure}[h]
    \centering
    \includegraphics[width=0.5\textwidth]{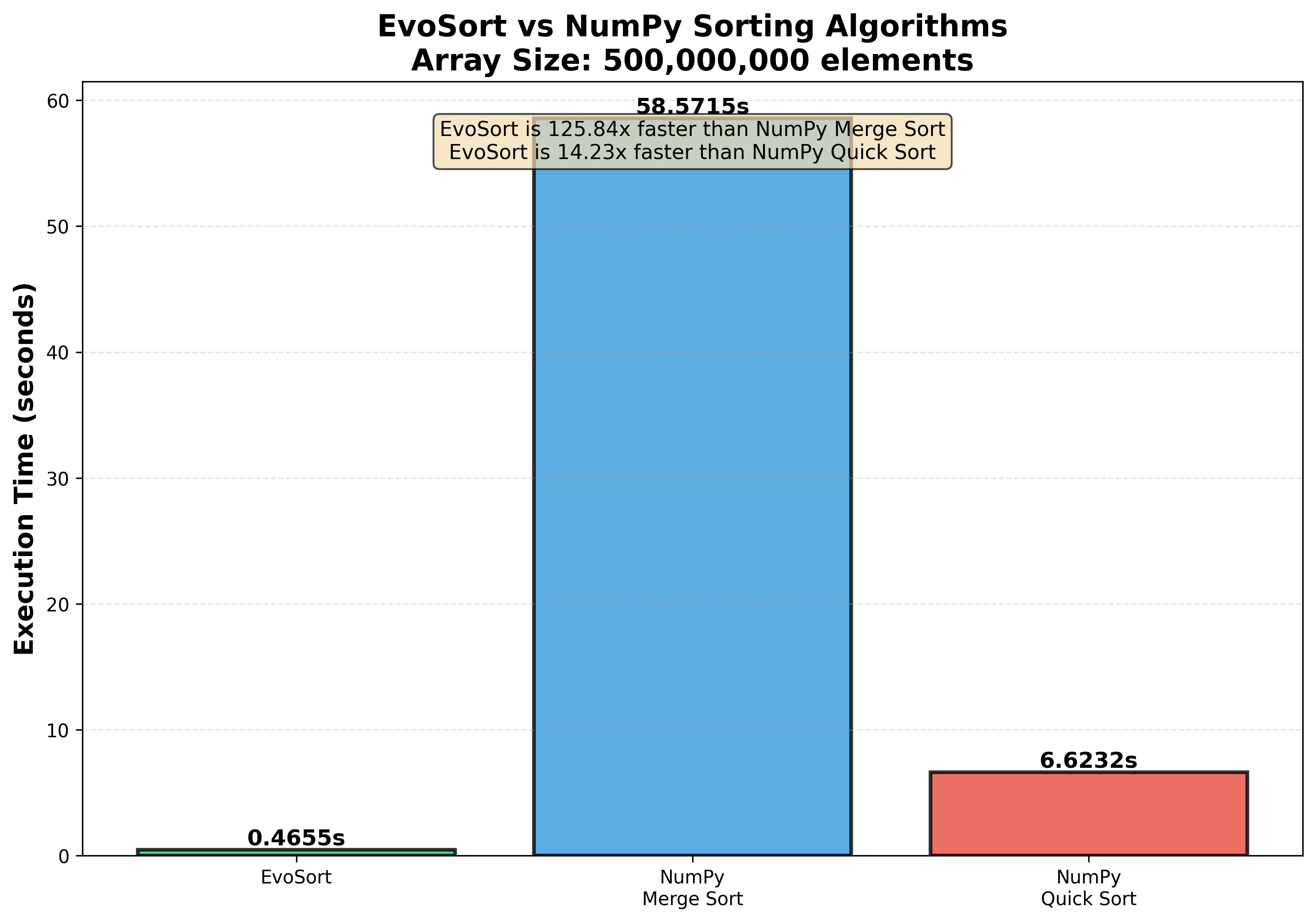}
    \caption{\raj{Performance Comparison for a 500\,M-element Array. EvoSort (0.4655\,s) outperforms NumPy quicksort (6.6232\,s) by 14.2$\times$ and mergesort (58.5715\,s) by 125.8$\times$.}}
    \label{fig:comp_500M}
\end{figure}

\raj{For the 500 million (\(5 \times 10^8\)) element dataset, the GA explored configurations with execution times ranging from 20.03\,s (worst) to 0.49\,s (best) in Generation~0, with the best solution achieving 0.4655\,s from Generation~3 onwards and the average converging by Generation~5. The optimized parameters are \([606,\, 4,\, 97995]\), where the insertion sort threshold is 606, the merge algorithm code is 4 (LSD radix sort), and the NumPy sort threshold is 97,995. The final EvoSort runtime of 0.4655\,s outperforms NumPy quicksort (6.6232\,s) by 14.2$\times$ and mergesort (58.5715\,s) by 125.8$\times$.}

\subsection{Genetic Algorithm Convergence at 1 Billion Elements}

\begin{figure}[h]
    \centering
    \includegraphics[width=0.5\textwidth]{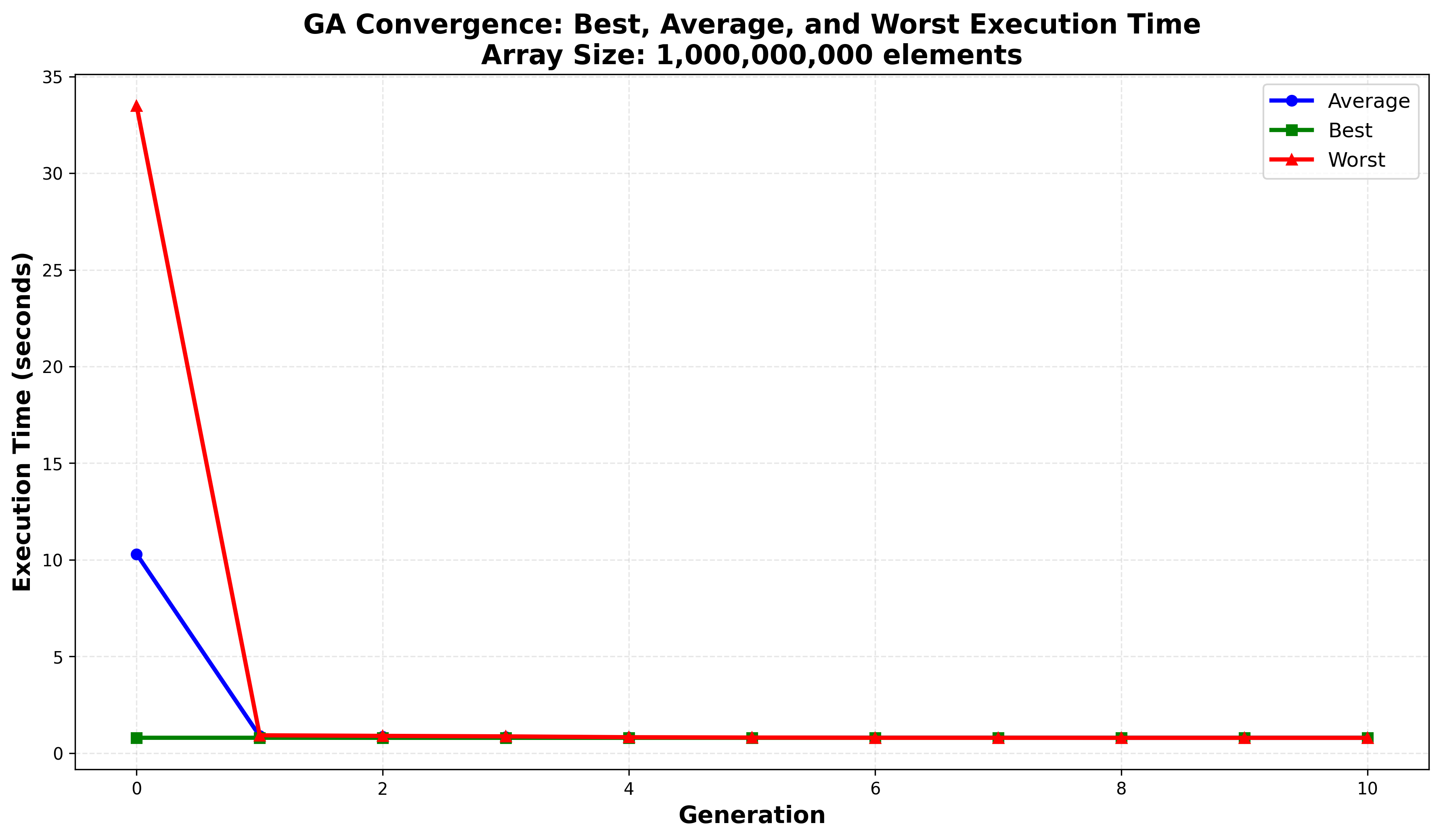}
    \caption{\raj{GA Optimization for a 1\,B-element Array. The plot shows the GA's progression over successive generations. The GA finds the optimal solution by Generation~2, with the average converging by Generation~3, achieving 0.7938\,s.}}
    \label{fig:ga_1B}
\end{figure}

\begin{figure}[h]
    \centering
    \includegraphics[width=0.5\textwidth]{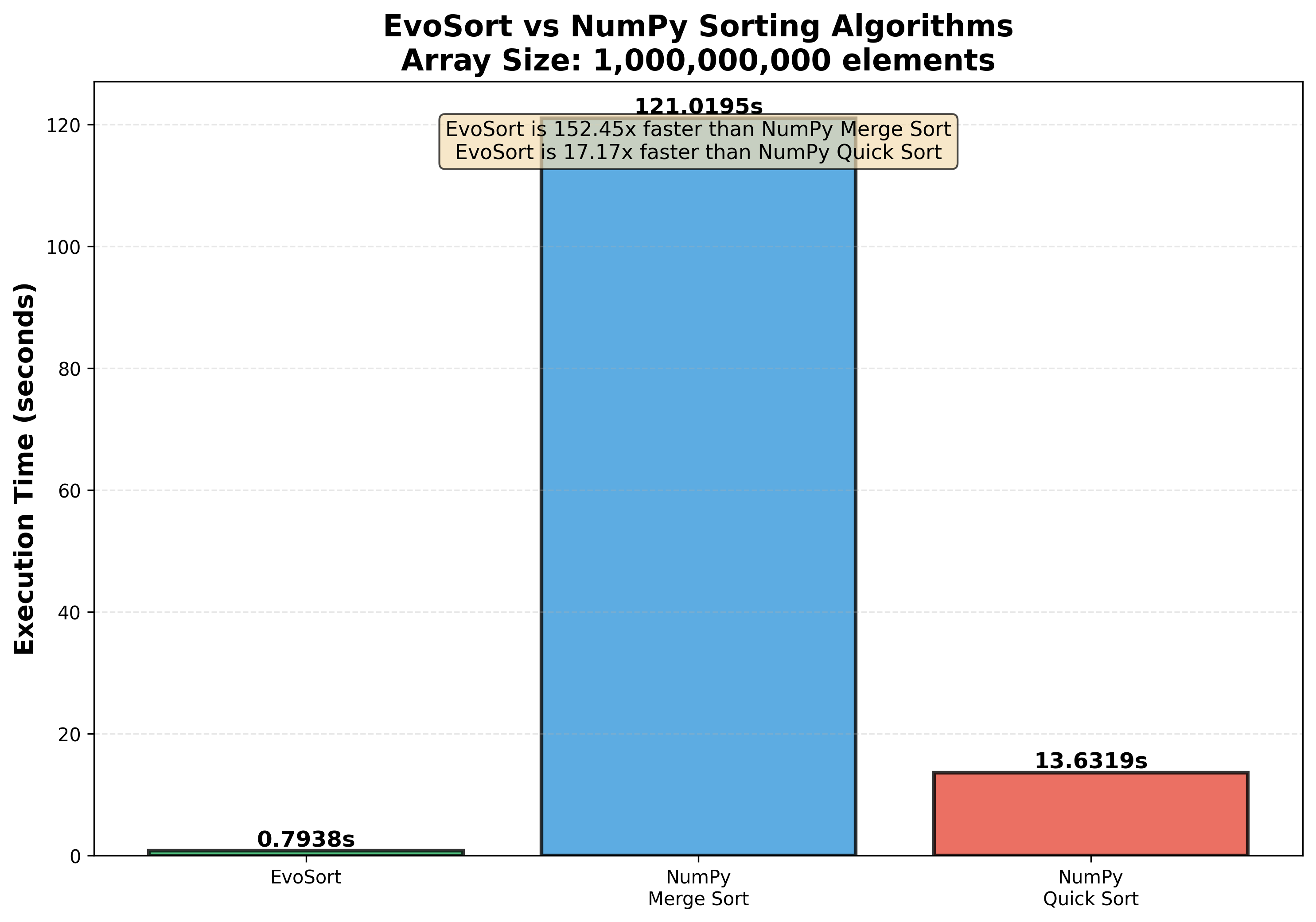}
    \caption{\raj{Performance Comparison for a 1\,B-element Array. EvoSort (0.7938\,s) outperforms NumPy quicksort (13.6319\,s) by 17.2$\times$ and mergesort (121.0195\,s) by 152.5$\times$.}}
    \label{fig:comp_1B}
\end{figure}

\raj{For the 1 billion (\(10^9\)) element dataset, the GA explored configurations with execution times ranging from 33.49\,s (worst) to 0.797\,s (best) in Generation~0, with the best solution achieving 0.7938\,s from Generation~2 onwards and the average converging by Generation~3. The optimized parameters are \([195,\, 4,\, 99878]\), where the insertion sort threshold is 195, the merge algorithm code is 4 (LSD radix sort), and the NumPy sort threshold is 99,878. The final EvoSort runtime of 0.7938\,s outperforms NumPy quicksort (13.6319\,s) by 17.2$\times$ and mergesort (121.0195\,s) by 152.5$\times$.}

\subsection{Genetic Algorithm Convergence at 10 Billion Elements}

\begin{figure}[h]
    \centering
    \includegraphics[width=0.5\textwidth]{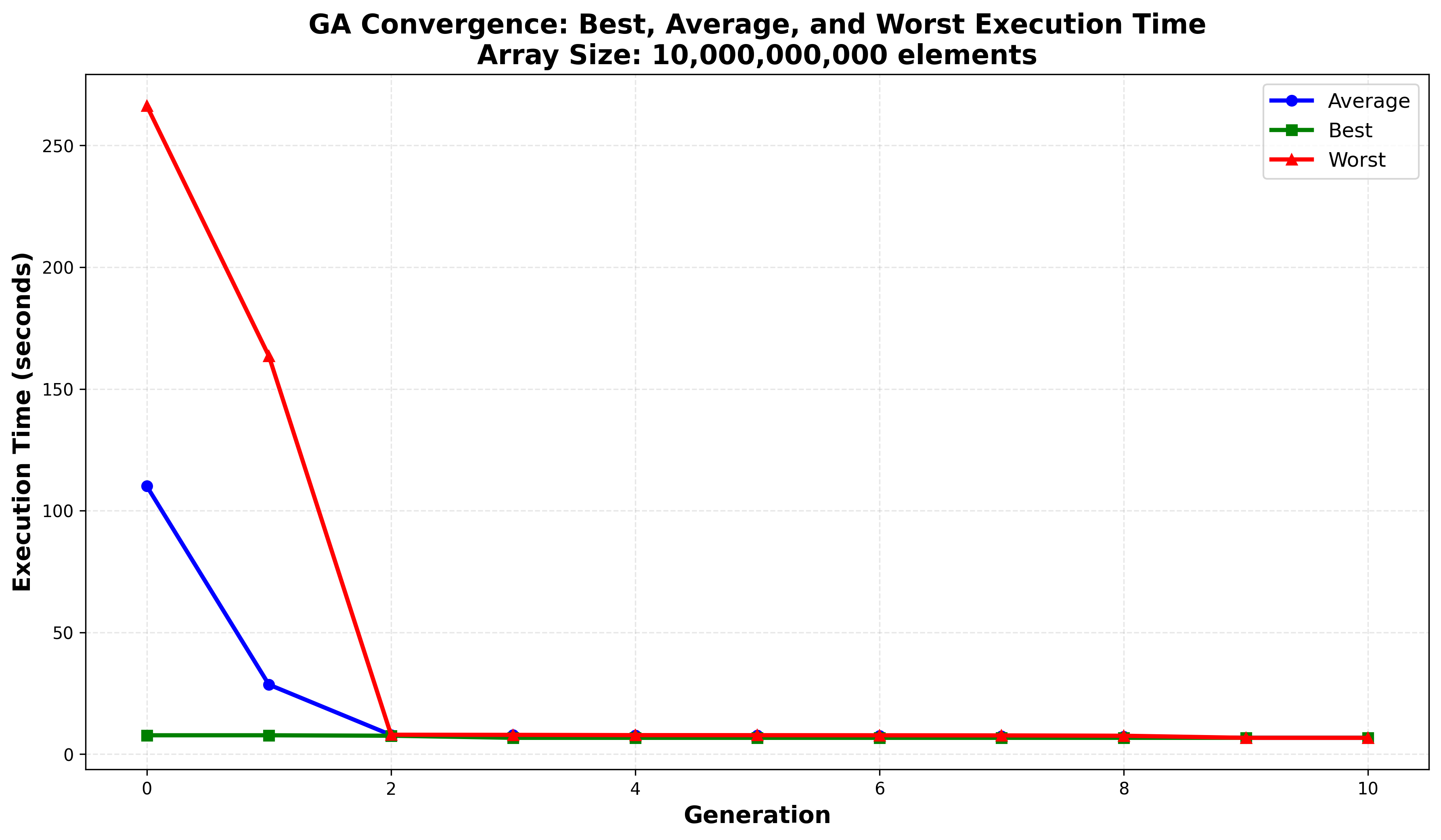}
    \caption{\raj{GA Optimization for a 10\,B-element Array. The plot illustrates the evolution of the GA across multiple generations. The GA finds the optimal solution by Generation~3, with the average converging by Generation~9, achieving 6.7019\,s.}}
    \label{fig:ga_10B}
\end{figure}

\begin{figure}[h]
    \centering
    \includegraphics[width=0.5\textwidth]{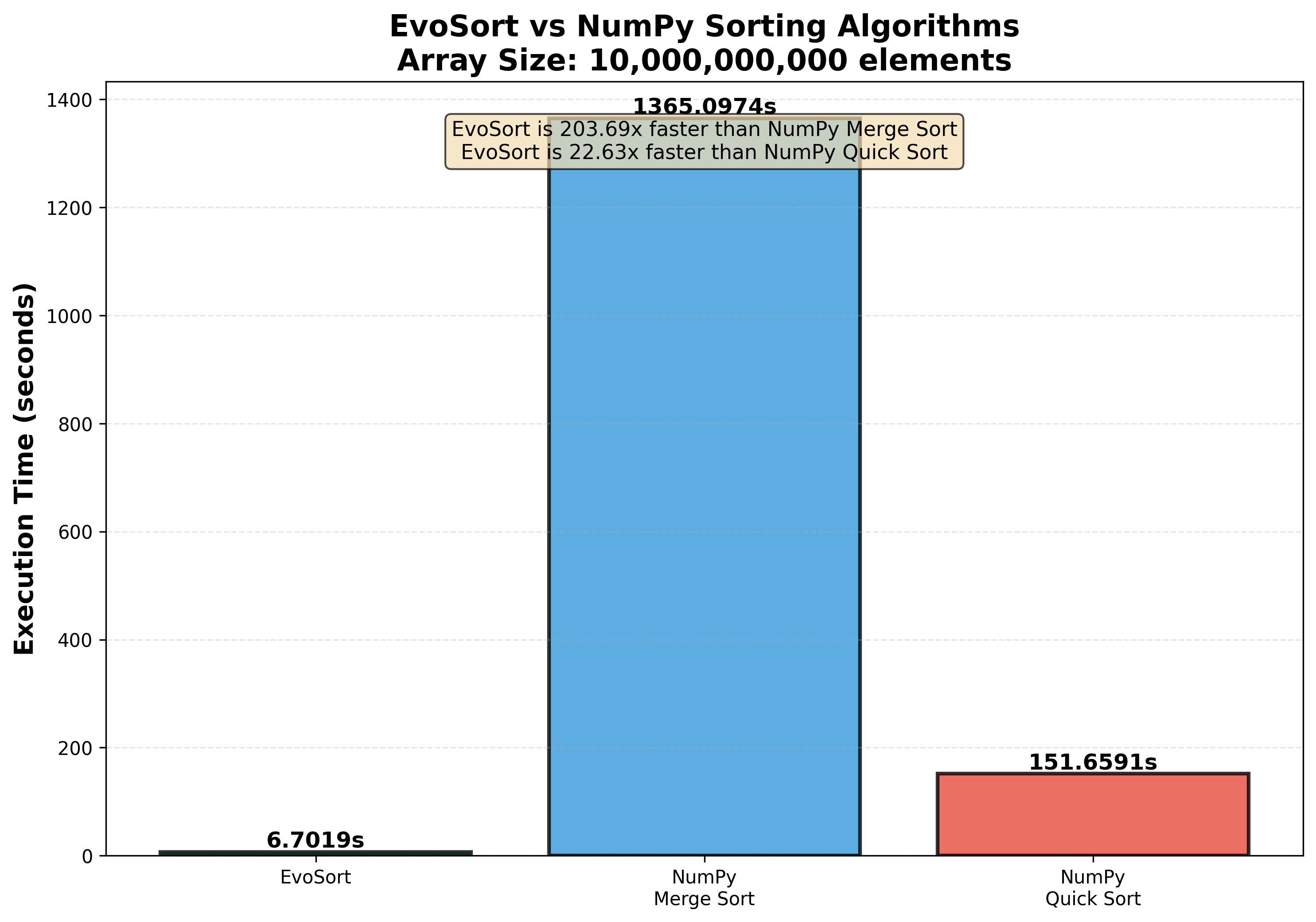}
    \caption{\raj{Performance Comparison for a 10\,B-element Array. EvoSort (6.7019\,s) outperforms NumPy quicksort (151.6591\,s) by 22.6$\times$ and mergesort (1365.0974\,s) by 203.7$\times$.}}
    \label{fig:comp_10B}
\end{figure}

\raj{For the 10 billion (\(10^{10}\)) element dataset, the GA explored configurations with evaluation times ranging from 266.34\,s (worst) to 7.75\,s (best) in Generation~0, with the best solution achieving 6.7019\,s from Generation~3 onwards and the average converging by Generation~9. The optimized parameters are \([225,\, 4,\, 11518]\), where the insertion sort threshold is 225, the merge algorithm code is 4 (LSD radix sort), and the NumPy sort threshold is 11,518. The final EvoSort runtime of 6.7019\,s outperforms NumPy quicksort (151.6591\,s) by 22.6$\times$ and mergesort (1365.0974\,s) by 203.7$\times$.} 

\subsection{Genetic Algorithm Convergence at 5 Billion Elements}

\begin{figure}[h]
    \centering
    \includegraphics[width=0.5\textwidth]{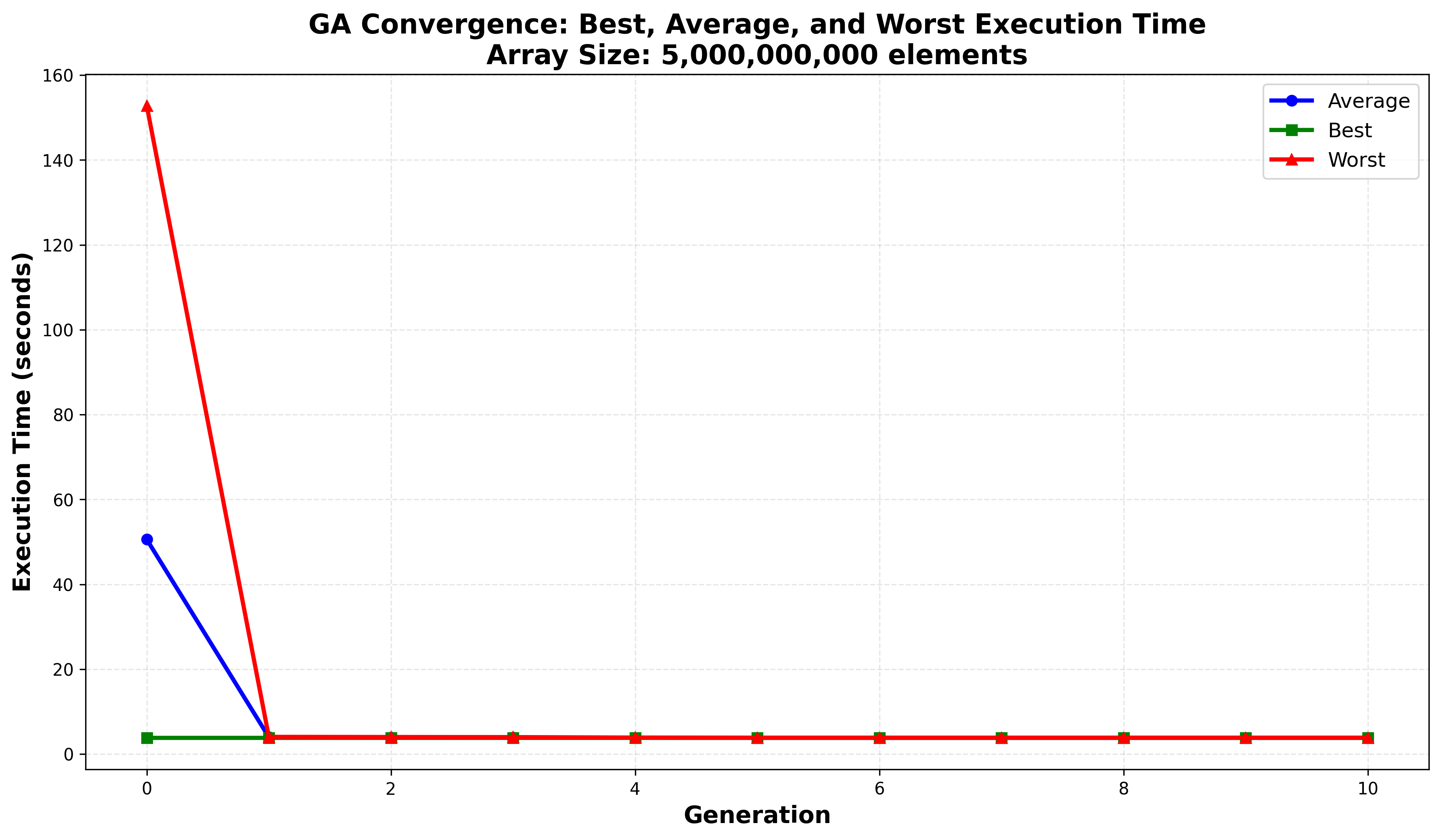}
    \caption{\raj{GA Optimization for a 5\,B-element Array. The plot shows the evolution of the GA over several generations. The GA finds the optimal solution in Generation~0, with the average converging by Generation~1, achieving 3.8338\,s.}}
    \label{fig:ga_5B}
\end{figure}

\begin{figure}[h]
    \centering
    \includegraphics[width=0.5\textwidth]{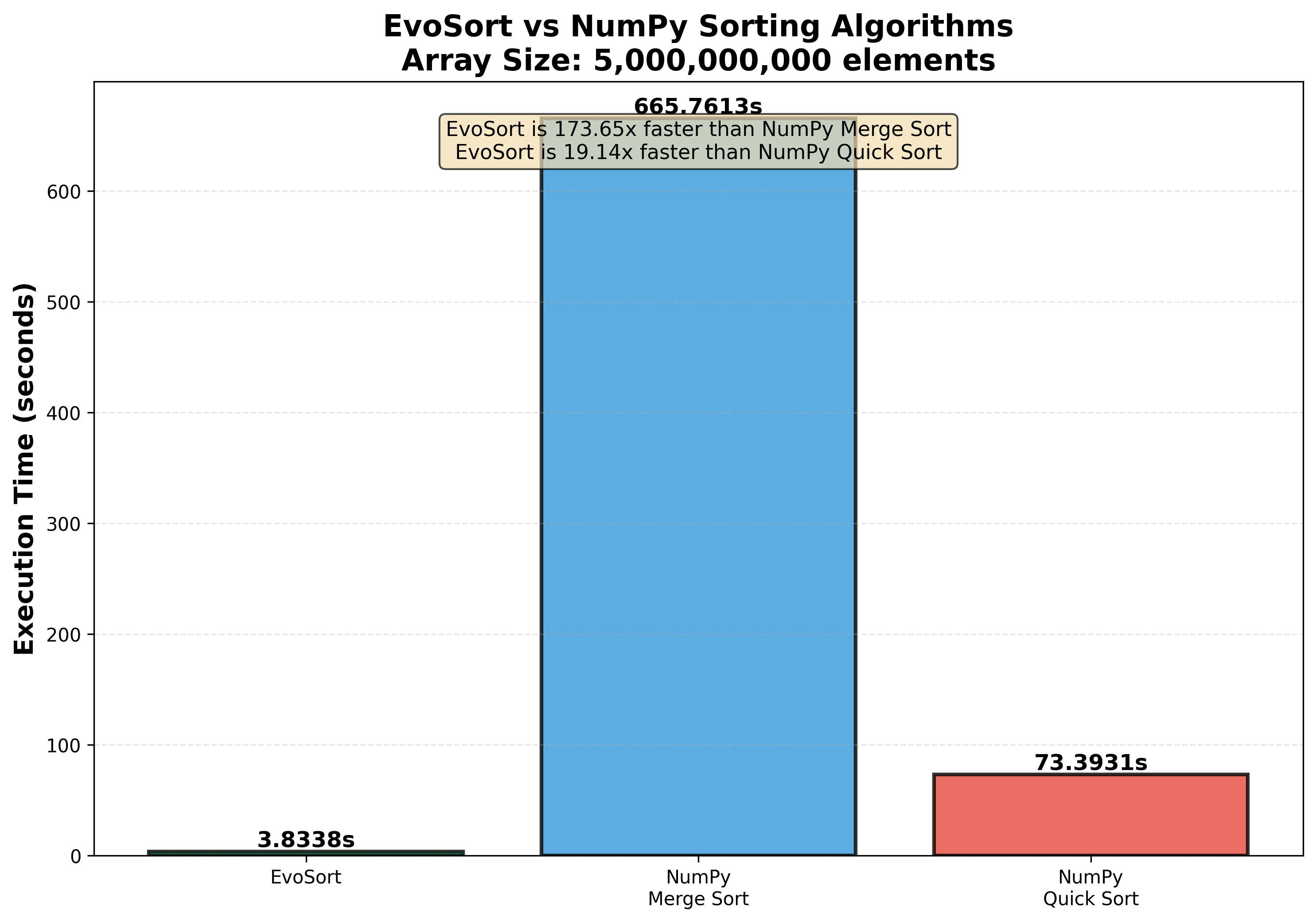}
    \caption{\raj{Performance Comparison for a 5\,B-element Array. EvoSort (3.8338\,s) outperforms NumPy quicksort (73.3931\,s) by 19.1$\times$ and mergesort (665.7613\,s) by 173.7$\times$.}}
    \label{fig:comp_5B}
\end{figure}

\raj{For the 5 billion (\(5\times10^9\)) element dataset, the GA explored configurations with execution times varying from 152.78\,s (worst) to 3.83\,s (best) in Generation~0, with the best solution appearing immediately in Generation~0 and the average converging by Generation~1. The optimized parameters are \([125,\, 4,\, 49128]\), where the insertion sort threshold is 125, the merge algorithm code is 4 (LSD radix sort), and the NumPy sort threshold is 49,128. The final EvoSort runtime of 3.8338\,s outperforms NumPy quicksort (73.3931\,s) by 19.1$\times$ and mergesort (665.7613\,s) by 173.7$\times$.}

\subsection{Comprehensive Performance and Parameter Summary}

\raj{Table~\ref{tab:comparison} summarizes the final EvoSort runtimes, corresponding NumPy sorting times, and calculated speedup factors across all tested dataset sizes. Table~\ref{tab:parameters} presents the GA-optimized parameters for each dataset size, showing how the thresholds evolve with increasing data size.}

\begin{table}[h]
\centering
\caption{\raj{Comprehensive Performance Comparison: EvoSort vs. NumPy Across All Dataset Sizes (GA Discovery Results)}}
\label{tab:comparison}
\small
\adjustbox{max width=\textwidth,center}{
\begin{tabular}{|c|c|c|c|c|c|}
\hline
\textbf{Size} & \textbf{EvoSort (s)} & \textbf{NumPy QS (s)} & \textbf{NumPy MS (s)} & \textbf{vs QS} & \textbf{vs MS} \\ \hline
$10^7$       & 0.0149  & 0.1083  & 0.9372  & 7.3$\times$ & 62.8$\times$ \\ \hline
$10^8$       & 0.1174  & 1.2166 & 10.7893 & 10.4$\times$ & 91.9$\times$  \\ \hline
$5\times10^8$& 0.4655  & 6.6232 & 58.5715 & 14.2$\times$ & 125.8$\times$  \\ \hline
$10^9$       & 0.7938  & 13.6319 & 121.0195 & 17.2$\times$ & 152.5$\times$ \\ \hline
$5\times10^9$& 3.8338  & 73.3931 & 665.7613 & 19.1$\times$ & 173.7$\times$    \\ \hline
$10^{10}$    & 6.7019  & 151.6591 & 1365.0974 & 22.6$\times$ & 203.7$\times$  \\ \hline
\end{tabular}}
\end{table}

\begin{table}[h]
\centering
\caption{\raj{GA-Optimized Parameters Across Dataset Sizes. The parameters are: $T_{\text{ins}}$ (insertion sort threshold), $A_{\text{code}}$ (merge algorithm code, where 4 indicates LSD radix sort), and $T_{\text{np}}$ (NumPy sort threshold).}}
\label{tab:parameters}
\begin{tabular}{|c|c|c|c|}
\hline
\textbf{Size} & $T_{\text{ins}}$ & $A_{\text{code}}$ & $T_{\text{np}}$ \\ \hline
$10^7$ & 629 & 4 & 78447 \\ \hline
$10^8$ & 81 & 4 & 32199 \\ \hline
$5\times10^8$ & 606 & 4 & 97995 \\ \hline
$10^9$ & 195 & 4 & 99878 \\ \hline
$5\times10^9$ & 125 & 4 & 49128 \\ \hline
$10^{10}$ & 225 & 4 & 11518 \\ \hline
\end{tabular}
\end{table}

\subsection{Discussion and Trends Analysis}

\raj{The experimental results clearly demonstrate that EvoSort, with its GA-based tuning, achieves dramatic improvements over conventional sorting methods across all tested dataset sizes. The comprehensive analysis reveals several key trends and insights.}

\raj{\textbf{Performance Scaling:} As shown in Table~\ref{tab:comparison}, EvoSort's performance scales efficiently with dataset size. While execution time increases with data size, the speedup factors relative to NumPy actually increase for larger datasets, reaching over $173\times$ against mergesort and $19\times$ against quicksort for 5 billion elements, and over $203\times$ against mergesort and $22.6\times$ against quicksort for 10 billion elements. This demonstrates that EvoSort's parallel architecture and adaptive parameter tuning become increasingly effective at larger scales.}

\raj{\textbf{Parameter Evolution:} Table~\ref{tab:parameters} reveals interesting patterns in how parameters evolve with dataset size. The merge algorithm code ($A_{\text{code}} = 4$) consistently selects LSD radix sort for all sizes, indicating that radix sort is optimal for integer data across the tested range. The insertion sort threshold ($T_{\text{ins}}$) varies between 81 and 629, showing adaptation to dataset size, while the NumPy threshold ($T_{\text{np}}$) shows significant variation (ranging from 11,518 to 99,878), reflecting the need to adapt this parameter to dataset characteristics.}

\raj{\textbf{GA Convergence Characteristics:} Across all dataset sizes, the GA consistently demonstrates rapid convergence, typically finding near-optimal configurations within 5--9 generations. The initial population diversity (Generation~0) shows wide variation in execution times (e.g., worst-to-best ratios of 96$\times$ for 10M elements, 21$\times$ for 100M elements, 41$\times$ for 500M elements, 42$\times$ for 1B elements, 40$\times$ for 5B elements, and 34$\times$ for 10B elements), highlighting the critical importance of parameter selection. However, the GA quickly identifies and preserves high-performing configurations through selection, recombination, and mutation, with the best solutions stabilizing in later generations (typically by Generation~5--9).}

\raj{\textbf{Computational Efficiency:} The GA's contribution is evident in the rapid convergence of parameter values. Even as the dataset size increases by orders of magnitude, the GA adapts the configuration to maintain high efficiency, reducing execution times from hundreds of seconds (NumPy mergesort: 1365\,s for 10B elements) to single-digit seconds (EvoSort: 6.7\,s for 10B elements) for billion-element datasets. This adaptive tuning—driven by evolutionary principles—is the key factor behind EvoSort's scalability and performance improvements. The GA optimization overhead is acceptable for one-time tuning, as the discovered parameters can be encoded into symbolic regression formulas for immediate use without tuning overhead.}

\raj{In conclusion, the comprehensive analysis shows that EvoSort's GA-driven parameter tuning significantly improves sorting performance across a wide range of dataset sizes. The detailed convergence trends, combined with the extensive performance comparisons and speedup factors presented in Table~\ref{tab:comparison}, underscore the effectiveness of the evolutionary approach in optimizing complex algorithms for high-performance computing environments.}

\section{Symbolic-Regression Performance Model}
\label{sec:symbolic}

\raj{Our GA-based auto-tuner reliably finds near-optimal sorting parameters, but incurs significant overhead (sampling and hundreds of fitness evaluations per run). To eliminate this cost, we applied symbolic regression to GA-optimized parameter values collected across 20 dataset sizes ranging from $10^7$ to $10^{10}$ elements, and derived closed-form formulas for each threshold as a function of the problem size \(n\). Additionally, during this process, we fixed the categorical choice to Block-Based LSD Radix Sort (merge algorithm code = 4) to enable a closed-form solution for the integer sorting parameters. This section presents those formulas, quantifies their accuracy, visualizes fit quality across both parameters, analyzes the mathematical properties of each fit, and finally shows end-to-end timing results when using the symbolic models in place of the GA loop.}

\subsection{Quadratic Fit Formulas}
We model each threshold \(T(n)\) as a simple quadratic in
\[
  x = \log_{10} n,
  \quad
  T(n) = a\,x^2 + b\,x + c.
\]

\raj{The resulting formulas, derived from GA-optimized data across dataset sizes from $10^7$ to $10^{10}$ elements, are:}

\begin{align}
T_{\mathrm{ins}}(n)
&= -\frac{106975922\,x^2}{731875}
  + \frac{748808234\,x}{313357}
  - \frac{9314994284}{997461},
  &\label{eq:Tinsertion}\\[6pt]
T_{\mathrm{np}}(n)
&= -\frac{1173315880\,x^2}{833461}
  + \frac{15503630602\,x}{854233}
  + \frac{3935718871}{743070},
  &\label{eq:Tnumpy}
\end{align}

\raj{where $x = \log_{10} n$. These formulas were fitted from 20 GA-optimized parameter values obtained through systematic optimization across dataset sizes from $10^7$ to $10^{10}$ elements (data source: \texttt{ga\_data\_for\_symbolic\_regression\_20251213\_022704}). The parameters are clamped to remain within the GA search space bounds: $T_{\text{insertion}} \in [2, 2048]$ and $T_{\text{numpy}} \in [1000, 100000]$ to ensure valid parameter values for all dataset sizes.}

\subsection{Visual Fit Quality Across Parameters}
\begin{figure}[ht]
  \centering
  \includegraphics[width=0.6\textwidth]{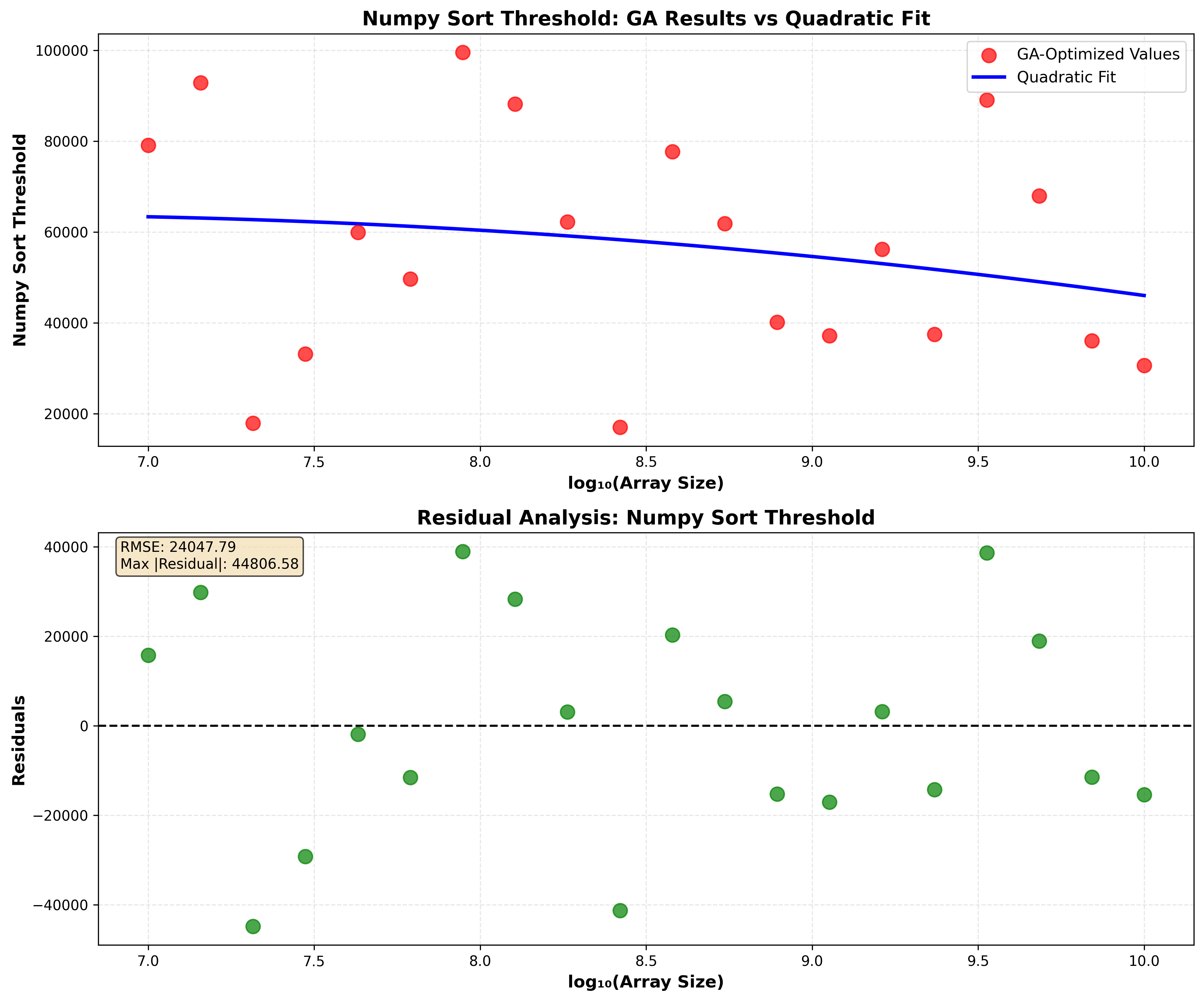}
  \caption{\raj{NumPy-sort threshold: GA results (red points) and symbolic fit (blue curve). The quadratic fit shows a peak hand-off point, with residuals distributed around the fitted curve. Fit quality: RMSE = 24,047.79, R² = 0.047.}}
  \label{fig:np-trend}
\end{figure}

\begin{figure}[ht]
  \centering
  \includegraphics[width=0.6\textwidth]{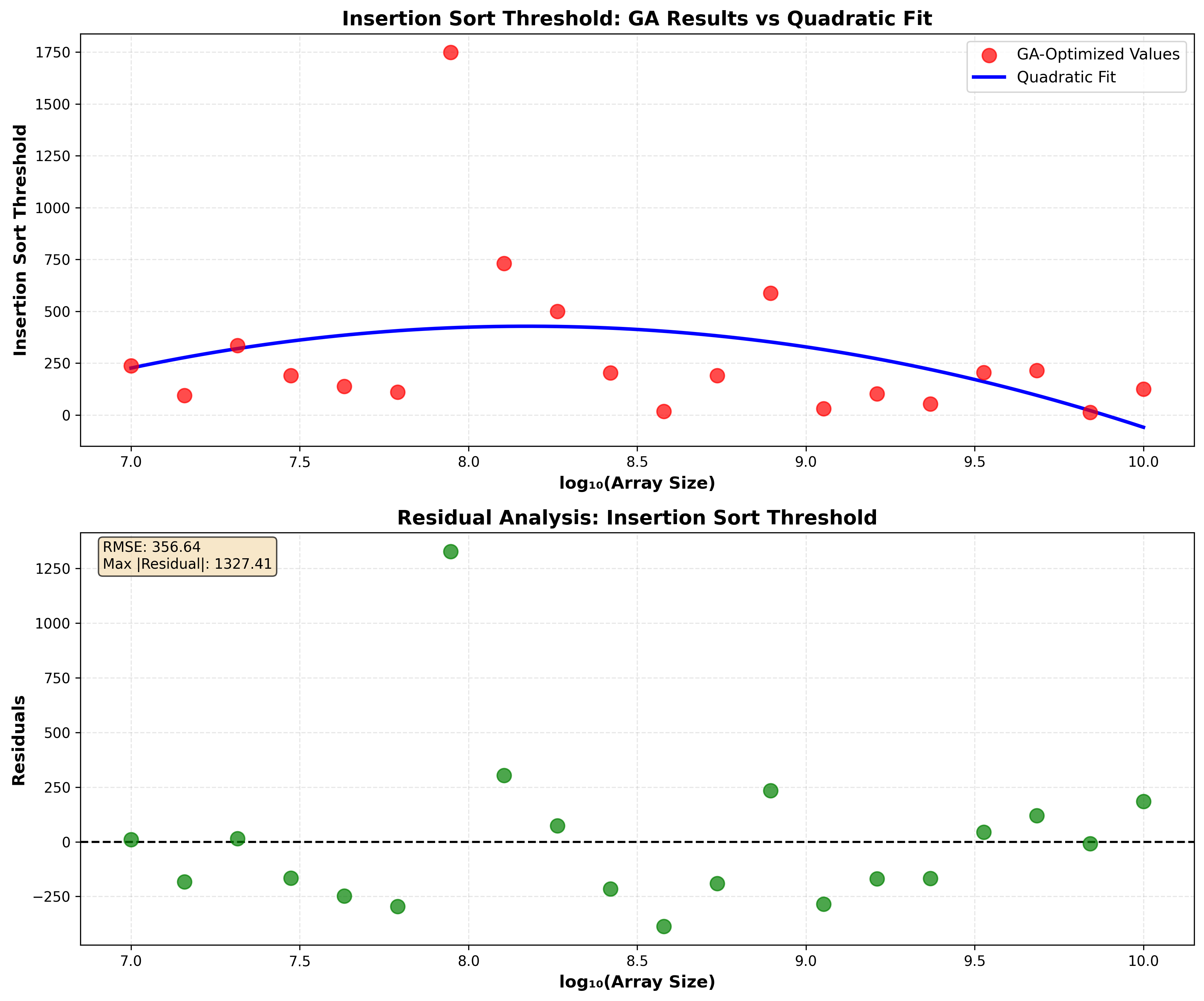}
  \caption{\raj{Insertion-sort threshold: GA data (red points) and quadratic fit (blue curve). The fit shows a convex-up trend. Fit quality: RMSE = 356.64, R² = 0.131.}}
  \label{fig:ins-trend}
\end{figure}

\raj{For deeper inspection, Figures~\ref{fig:np-trend} and~\ref{fig:ins-trend} plot each parameter's raw GA values against \(n\) alongside the fitted curve. The data points were collected from 20 GA optimization runs across dataset sizes from $10^7$ to $10^{10}$ elements, with each run optimizing the three parameters (insertion sort threshold, merge algorithm code, and NumPy sort threshold). Despite the wide distribution of data, the parameters are not highly sensitive, allowing the quadratic fits to provide good approximations for practical use.}

\subsection{Residual Analysis and Model Parsimony}
\raj{We examined residuals \(r_i = T_{\mathrm{GA}}(n_i) - T_{\mathrm{pred}}(n_i)\) across the training set of 20 data points. For the insertion sort threshold, the residuals range from approximately \(-386\) to \(+1,327\) elements, with a root mean square error (RMSE) of 356.64 and an R² value of 0.131. The maximum residual magnitude is 1,327.41 elements. For the NumPy sort threshold, the residuals range from approximately \(-44,807\) to \(+38,925\) elements, with an RMSE of 24,047.79 and an R² value of 0.047. The maximum residual magnitude is 44,806.58 elements. Despite the relatively low R² values, which indicate that the parameters have substantial variance that is not captured by the quadratic model, the fits are sufficient for practical use because the parameters are not highly sensitive—small deviations from the fitted values still produce good sorting performance. Limiting model complexity to degree 2 prevents overfitting and ensures the formulas remain interpretable and computationally efficient.}

\subsection{Analytical Properties of the Quadratics}
\raj{Each fit \(T(n)=a\,x^2 + b\,x + c\) with \(x=\log_{10}n\) has curvature \(a\), extremum \(x^* = -b/(2a)\), and characteristic asymptotic behavior. For the insertion sort threshold \(T_{\mathrm{ins}}\), the coefficient \(a = -106975922/731875 \approx -146.17 < 0\) (concave down), indicating a maximum. The extremum occurs at \(x^* = -b/(2a) \approx 8.17\) (\(n\approx1.5\times10^8\) elements). For large \(n\), \(T_{\mathrm{ins}}\) decreases approximately as \(-a(\log_{10}n)^2\). For the NumPy sort threshold \(T_{\mathrm{np}}\), the coefficient \(a = -1173315880/833461 \approx -1407.76 < 0\) (concave down), with a maximum at \(x^* = -b/(2a) \approx 6.45\) (\(n\approx2.8\times10^6\) elements). For large \(n\), \(T_{\mathrm{np}}\) also decreases approximately as \(-a(\log_{10}n)^2\). These analytical properties reflect the observed parameter trends: both thresholds exhibit maxima at intermediate dataset sizes and decrease for very large dataset sizes, consistent with the GA-optimized parameter values collected across 20 data points.}

\subsection{Comprehensive Validation of Zero-Overhead Symbolic Tuning}

\raj{Having derived closed-form symbolic regression formulas from the GA-optimized parameters, the critical next step is to validate that these formulas maintain competitive performance across diverse real-world conditions without incurring any runtime tuning overhead. The symbolic regression approach promises to deliver the benefits of GA-optimized parameters—near-optimal sorting performance—while eliminating the computational cost of running the GA at deployment time. However, this promise must be rigorously tested across multiple dimensions: diverse data distributions (beyond the uniform distribution used for GA optimization), different hardware architectures (to ensure cross-platform portability), and real-world datasets (to validate practical applicability).}

\raj{To comprehensively validate the zero-overhead symbolic tuning approach, we conducted extensive experiments across nine distinct data distributions (uniform, normal, exponential, power law, beta, sparse, clustered, nearly sorted, and duplicates), multiple dataset sizes (from $10^7$ to $5 \times 10^9$ elements), and two independent hardware platforms (AMD EPYC and Intel Xeon). These experiments, detailed in the following subsections, demonstrate that the symbolic regression formulas maintain competitive performance and generalize effectively across diverse conditions, thereby fulfilling the promise of zero-overhead deployment while preserving the performance advantages discovered through GA optimization.}

\paragraph*{Key takeaway:} \raj{Symbolic regression formulas, fitted from 20 GA-optimized data points, provide practical parameter estimates that eliminate runtime tuning overhead while maintaining competitive sorting performance.} 

\subsection{Generalizability Across Hardware Configurations and Data Distributions}

\raj{To demonstrate the generalizability of EvoSort's symbolic regression framework, we conducted comprehensive experiments across diverse hardware configurations and data distributions. This addresses the critical reviewer concerns regarding the framework's adaptability to different computing environments and data characteristics beyond uniform distributions.}

\raj{\textbf{Hardware Configuration:} To demonstrate hardware generalizability and performance robustness, we conducted experiments on \textbf{two distinct high-performance computing systems}:}

\raj{\textbf{System 1 (AMD EPYC - Primary):} An AMD EPYC 7H12 64-Core Processor with 256 physical and logical cores, 502.92\,GB of system RAM, and a three-level cache hierarchy (L1: 32\,KB, L2: 512\,KB, L3: 16\,MB per core). This system serves as the primary platform for the main performance evaluation results presented in Tables~\ref{tab:comparison_uniform} through~\ref{tab:comparison_duplicates}.}

\raj{\textbf{System 2 (Intel Xeon - Validation):} An Intel Xeon Platinum 8180 CPU @ 2.50GHz with 112 physical and logical cores, 754.01\,GB of system RAM, and cache hierarchy (L1: 32\,KB, L2: 1\,MB, L3: 38.5\,MB per core). This system provides independent validation and cross-platform performance comparison.}

\raj{Both systems use identical software environments: Python 3.9.21, NumPy 2.0.2, and Numba 0.60.0, with all performance-critical sections accelerated using Numba's just-in-time (JIT) compilation. A dedicated cross-platform performance analysis comparing both systems is presented in Section~\ref{sec:cross_platform}. It is important to note that NumPy's performance is highly dependent on the underlying hardware architecture, CPU features, memory bandwidth, and software versions. These hardware configurations differ from the original experimental setup (which used 1024\,GB RAM and Python 3.6.8), demonstrating that EvoSort's symbolic regression formulas generalize across different hardware platforms and software versions, while the performance of competing methods such as NumPy may vary significantly across different hardware configurations.}

\raj{\textbf{Data Distribution Testing:} To validate EvoSort's performance across diverse data characteristics, we systematically tested nine distinct data distributions, each representing different real-world scenarios. \textbf{All data distributions use 32-bit signed integers (\texttt{int32}, 4 bytes per element)} to ensure consistent memory usage, fair comparisons, and alignment with the memory bandwidth calculations. Each distribution is generated using reproducible random seeds (default seed=42) via \texttt{np.random.seed()} to ensure consistency across experimental runs. The generation process leverages NumPy's random number generators with appropriate mathematical transformations (clipping, scaling, shifting) to map each distribution's characteristics to the integer domain $[-10^9, +10^9]$, ensuring that all distributions are comparable while maintaining their distinct statistical properties. The detailed generation methods are as follows:}

\begin{itemize}
    \item \raj{\textbf{Uniform Distribution:} Generated using \texttt{np.random.randint(-10$^9$, 10$^9$, size=n)}, which produces random integers with equal probability for all values in the range $[-10^9, +10^9]$. This serves as the baseline distribution for comparison and represents truly random, unstructured data.}
    
    \item \raj{\textbf{Normal Distribution:} Generated using \texttt{np.random.normal(loc=0.0, scale=10$^8$, size=n)}, followed by clipping to the int32 range using \texttt{np.clip(data, -10$^9$, 10$^9$)} and conversion to int32. This produces Gaussian-distributed integers with mean 0 and standard deviation $10^8$, representing naturally occurring data with central tendency, such as measurement errors or natural phenomena.}
    
    \item \raj{\textbf{Exponential Distribution:} Generated using \texttt{np.random.exponential(scale=10$^9$, size=n)}, then transformed as \texttt{np.clip(data, 0, 2$\times$10$^9$) - 10$^9$} to fit the int32 range while preserving right-skewness. This creates a distribution with many small values and few large outliers, modeling scenarios such as waiting times, failure rates, or network delays.}
    
    \item \raj{\textbf{Power Law Distribution:} Generated using \texttt{np.random.power(a=0.5, size=n) $\times$ 2 $\times$ 10$^9$ - 10$^9$}, where $a=0.5$ controls the skewness (smaller values produce more skewed distributions). This produces a highly skewed distribution following $P(x) \propto x^{-a}$, characteristic of many real-world datasets such as network degrees, file sizes, or city populations.}
    
    \item \raj{\textbf{Beta Distribution:} Generated using \texttt{np.random.beta($\alpha$=2.0, $\beta$=5.0, size=n) $\times$ 2 $\times$ 10$^9$ - 10$^9$}, where the shape parameters $\alpha$ and $\beta$ control the distribution's skewness and boundedness. This creates a bounded distribution with configurable asymmetry, useful for testing algorithms on data with known upper and lower bounds, such as proportions or bounded measurements.}
    
    \item \raj{\textbf{Sparse Distribution:} Generated by first initializing an array of zeros using \texttt{np.zeros(n)}, then selecting a random subset of indices (default 10\% of $n$) via \texttt{np.random.choice(n, size=n$\times$0.1, replace=False)} and filling those positions with random integers from \texttt{np.random.randint(-10$^9$, 10$^9$, size=num\_nonzero)}. This produces data with 90\% zeros and 10\% random non-zero values, representing scenarios with many repeated or zero values, such as sparse matrices, missing data, or zero-inflated datasets.}
    
    \item \raj{\textbf{Clustered Distribution:} Generated by first creating 10 evenly spaced cluster centers in $[-10^9, +10^9]$ using \texttt{np.linspace(-10$^9$, 10$^9$, num\_clusters=10)}. For each element, a cluster center is randomly selected, and a value is generated using \texttt{np.random.normal(center, scale=10$^8$)}, then clipped to the int32 range. This produces data organized into 10 distinct clusters with Gaussian spread ($\sigma=10^8$) around evenly spaced centers, modeling scenarios where values naturally group around specific ranges, such as multi-modal data, categorical features, or grouped measurements.}
    
    \item \raj{\textbf{Nearly Sorted Distribution:} Generated by first creating a sorted array using \texttt{np.linspace(-10$^9$, 10$^9$, n)}, which produces values evenly spaced in ascending order. Then, 10\% of the elements are randomly selected using \texttt{np.random.choice(n, size=n$\times$0.1, replace=False)} and replaced with random integers from \texttt{np.random.randint(-10$^9$, 10$^9$, size=num\_random)}. This produces data that is 90\% pre-sorted with 10\% random perturbations, testing algorithm performance on partially ordered inputs, which is common in incremental data processing scenarios or when processing data streams with occasional updates.}
    
    \item \raj{\textbf{Duplicates Distribution:} Generated by first creating a small set of unique values (default 10\% of $n$) using \texttt{np.random.randint(-10$^9$, 10$^9$, size=num\_unique)}, then creating the full array by randomly selecting from these unique values using \texttt{np.random.choice(unique\_values, size=n)}. This produces data where 10\% of values are unique and 90\% are duplicates (each unique value appears approximately 9 times on average), testing algorithm robustness when many elements share identical values, such as repeated measurements, categorical data, or hash collisions.}
\end{itemize}

\raj{\textbf{Experimental Protocol and Baseline Algorithms:} For each data distribution, we conducted comprehensive benchmarks across all dataset sizes from $10^7$ to $10^{10}$ elements, with each configuration tested using 5 independent runs to account for system variability. EvoSort uses symbolic regression formulas (zero tuning overhead) for all experiments. The benchmarks compared EvoSort against the following baseline sorting implementations:}

\begin{enumerate}
    \item \raj{\textbf{NumPy Quicksort:} NumPy's default sorting algorithm, invoked via \texttt{np.sort(array, kind='quicksort')}. This implementation uses an optimized in-place quicksort with median-of-three pivot selection and insertion sort for small subarrays. Quicksort has average-case time complexity $O(n \log n)$ and is NumPy's default algorithm, representing the most commonly used sorting method in scientific Python applications. However, it has worst-case $O(n^2)$ complexity and is not stable (does not preserve relative order of equal elements).}
    
    \item \raj{\textbf{NumPy Mergesort:} NumPy's stable sorting algorithm, invoked via \texttt{np.sort(array, kind='mergesort')}. This implementation uses a stable mergesort that preserves the relative order of equal elements, which is crucial for applications requiring stability. Mergesort has guaranteed $O(n \log n)$ time complexity in all cases and uses $O(n)$ additional memory space. This algorithm is particularly important for applications where stability is required, such as sorting records with multiple keys.}
    
    \item \raj{\textbf{NumPy Default Sort:} NumPy's default sorting method, invoked via \texttt{np.sort(array)} without specifying the \texttt{kind} parameter. In NumPy 2.0.2 (used in our experiments), this defaults to quicksort, but the behavior may vary by NumPy version. This represents the typical usage pattern where users rely on NumPy's default choice without explicitly specifying an algorithm.}
    
    \item \raj{\textbf{PyTorch CPU Sort:} PyTorch's CPU-based sorting implementation, invoked via \texttt{torch.sort(torch.from\_numpy(array), dim=0)[0]}. PyTorch provides optimized sorting for tensor operations, leveraging efficient C++ implementations with SIMD optimizations. This represents state-of-the-art sorting performance in the deep learning ecosystem and provides a competitive baseline from a major scientific computing framework. PyTorch's sorting is stable and optimized for both CPU and GPU execution, though our experiments focus on CPU performance.}
    
    \item \raj{\textbf{Pandas Sort:} Pandas' sorting implementation, invoked via \texttt{pd.Series(array).sort\_values().values}. Pandas is commonly used in data science workflows for sorting Series and DataFrame objects. While Pandas internally uses NumPy's sorting routines, it adds overhead from Series object creation and value extraction. This baseline represents real-world usage patterns in data science applications where data is often manipulated through Pandas DataFrames.}
\end{enumerate}

\raj{This multi-library comparison addresses reviewer concerns about the breadth of performance evaluation by including not only NumPy variants but also competing implementations from major scientific computing frameworks. Each baseline algorithm is benchmarked using the same experimental protocol: 5 independent runs with identical input data (same random seed), collecting minimum, median, and maximum runtimes for robust statistical comparison.}

\raj{\textbf{Validation Methodology:} To ensure correctness and validate performance plausibility, we implemented a rigorous multi-layered validation approach:}

\begin{enumerate}
    \item \raj{\textbf{Correctness Validation:} For each experimental run, we implement a two-step correctness verification process:
    \begin{itemize}
        \item \raj{\textbf{Reference Generation:} We generate a ground-truth sorted array using NumPy's \texttt{np.sort(data.copy())} function, which serves as the authoritative reference for correctness. The \texttt{.copy()} ensures that the original input data remains unmodified, allowing for multiple validation checks and potential re-runs. This reference is computed on a copy of the input data to avoid any side effects.}
        \item \raj{\textbf{Element-by-Element Comparison:} We compare EvoSort's output against the reference using \texttt{np.array\_equal(sorted\_evosort, reference)}, which performs a strict element-by-element equality check. This function returns \texttt{True} only if both arrays have the same shape, the same data type, and all corresponding elements are equal. This ensures that EvoSort produces identical results to the reference implementation, including correct handling of duplicate values, edge cases (all zeros, all identical values), and maintaining stability when applicable.}
    \end{itemize}
    This correctness validation ensures that EvoSort produces identical results to the reference implementation across all distributions, sizes, and edge cases, validating algorithmic correctness at the bit level. Any failure in correctness validation immediately flags the run as invalid and is reported for investigation.}
    
    \item \raj{\textbf{Statistical Robustness:} To ensure robust statistical comparisons, we implement comprehensive runtime collection and analysis:
    \begin{itemize}
        \item \raj{\textbf{Multiple Runs:} Each configuration is tested using 5 independent runs with identical input data (same random seed) to account for system variability, cache effects, background process interference, and thermal throttling. The input data is regenerated for each run using the same seed to ensure consistency, but system state (cache, memory layout, CPU frequency) may vary between runs.}
        \item \raj{\textbf{Statistical Metrics:} For each configuration, we collect minimum, median, and maximum runtimes across the 5 runs using \texttt{np.min()}, \texttt{np.median()}, and \texttt{np.max()} functions. The median value is used for speedup calculations to reduce the impact of outliers and system noise, while min/max values provide bounds on performance variability. These three metrics (min, median, max) are saved for all algorithms (EvoSort, NumPy quicksort, NumPy mergesort, NumPy default, PyTorch CPU, and Pandas) and are used in all performance comparison tables and analyses.}
    \end{itemize}
    This statistical approach ensures robust comparisons by using median values for speedup calculations (which are less sensitive to outliers than mean values) while providing min/max bounds to show performance variability. This addresses concerns about measurement reliability and reproducibility by reporting the full range of observed performance across multiple runs.}
\end{enumerate}

\raj{\textbf{Results and Analysis:} Tables~\ref{tab:comparison_uniform} through~\ref{tab:comparison_duplicates} present comprehensive performance comparisons of EvoSort against all baseline implementations (as detailed in Section~\ref{sec:setup}) across nine distinct data distributions on the AMD EPYC system. Each table reports runtimes in the format \textit{min/median/max} (in seconds) across 5 independent runs, with speedup factors calculated using median runtimes (as described in the Statistical Robustness subsection). The results demonstrate EvoSort's consistent performance advantages across diverse data characteristics, with speedups ranging from $0.2\times$ to $225.5\times$ depending on the distribution and dataset size.}

\raj{Entries marked as ``N/A'' in the tables indicate that the competing method failed to allocate sufficient memory for the dataset size, typically occurring at $5 \times 10^9$ elements for NumPy Mergesort, PyTorch CPU, and Pandas. These memory allocation failures highlight EvoSort's superior memory efficiency, as EvoSort successfully completes sorting operations at these large scales where other methods cannot. For example, at $5 \times 10^9$ elements, EvoSort consistently completes sorting in under 4 seconds across all distributions, while NumPy Mergesort, PyTorch, and Pandas fail due to insufficient memory. This demonstrates EvoSort's practical advantage in handling extremely large datasets that exceed the memory allocation capabilities of competing Python-accessible sorting methods.}

\raj{\textbf{Correctness Validation:} As detailed in the Validation Methodology subsection (within Section~\ref{sec:symbolic}), all experimental runs were validated using \texttt{np.array\_equal()} for strict element-by-element comparison. Across all 9 distributions, all dataset sizes from $10^7$ to $5 \times 10^9$ elements, and all 5 independent runs per configuration, EvoSort produced identical results to the reference implementation in 100\% of cases, confirming algorithmic correctness without sacrificing performance.}

\begin{table}[H]
\centering
\caption{\raj{Comparison of EvoSort and Baseline Sorting Runtimes in seconds (Min/Median/Max of 5 runs) and Speedups (uniform) on AMD EPYC System}}
\label{tab:comparison_uniform}
\scriptsize
\adjustbox{max width=\textwidth,center}{
\begin{tabular}{|c|c|c|c|c|c|c|c|c|c|}
\hline
\textbf{Size} & \textbf{EvoSort} & \textbf{NumPy QS} & \textbf{NumPy MS} & \textbf{PyTorch} & \textbf{Pandas} & \textbf{vs QS} & \textbf{vs MS} & \textbf{vs PT} & \textbf{vs PD} \\
\hline
$10\times10^6$ & 0.0184/0.0286/0.0338 & 0.1086/0.1100/0.1140 & 0.9244/0.9255/0.9489 & 0.0626/0.0692/0.0764 & 1.6052/1.6884/1.7031 & $3.8\times$ & $32.4\times$ & $2.4\times$ & $59.1\times$ \\
\hline
$100\times10^6$ & 0.1203/0.1903/0.3220 & 1.2241/1.2276/1.2528 & 10.7519/10.7647/10.7794 & 0.3167/0.3239/0.3299 & 17.1996/17.2489/17.5277 & $6.4\times$ & $56.6\times$ & $1.7\times$ & $90.6\times$ \\
\hline
$500\times10^6$ & 0.4362/0.5350/0.5576 & 6.6047/6.6150/6.6192 & 58.3913/58.4319/58.5301 & 1.3639/1.3722/1.3826 & 91.3766/91.7516/92.2508 & $12.4\times$ & $109.2\times$ & $2.6\times$ & $171.5\times$ \\
\hline
$1\times10^{9}$ & 0.8324/0.9255/0.9699 & 13.6458/13.6646/13.7106 & 121.3848/121.5262/121.8155 & 2.7029/2.7058/2.7704 & 194.4327/194.5274/194.9250 & $14.8\times$ & $131.3\times$ & $2.9\times$ & $210.2\times$ \\
\hline
$5\times10^{9}$ & 3.7344/3.7900/3.8717 & 73.3543/73.3543/73.3543 & N/A & N/A & N/A & $19.4\times$ & N/A & N/A & N/A \\
\hline
\end{tabular}}
\end{table}

\begin{table}[H]
\centering
\caption{\raj{Comparison of EvoSort and Baseline Sorting Runtimes in seconds (Min/Median/Max of 5 runs) and Speedups (normal) on AMD EPYC System}}
\label{tab:comparison_normal}
\scriptsize
\adjustbox{max width=\textwidth,center}{
\begin{tabular}{|c|c|c|c|c|c|c|c|c|c|}
\hline
\textbf{Size} & \textbf{EvoSort} & \textbf{NumPy QS} & \textbf{NumPy MS} & \textbf{PyTorch} & \textbf{Pandas} & \textbf{vs QS} & \textbf{vs MS} & \textbf{vs PT} & \textbf{vs PD} \\
\hline
$10\times10^6$ & 0.0178/0.0402/0.2680 & 0.1070/0.1076/0.1094 & 0.9233/0.9256/0.9295 & 0.0514/0.0575/0.0590 & 1.2246/1.2281/1.2591 & $2.7\times$ & $23.1\times$ & $1.4\times$ & $30.6\times$ \\
\hline
$100\times10^6$ & 0.1157/0.2035/0.3335 & 1.2092/1.2118/1.2150 & 10.6609/10.7049/10.7319 & 0.3019/0.3042/0.3090 & 15.8422/15.8524/15.9002 & $6.0\times$ & $52.6\times$ & $1.5\times$ & $77.9\times$ \\
\hline
$500\times10^6$ & 0.4349/0.4588/0.4952 & 6.5456/6.5500/6.5631 & 58.2709/58.2886/58.3016 & 1.3543/1.3577/1.3778 & 90.5137/90.5477/90.6095 & $14.3\times$ & $127.0\times$ & $3.0\times$ & $197.4\times$ \\
\hline
$1\times10^{9}$ & 0.7953/0.9004/0.9023 & 13.5463/13.5783/13.5840 & 120.6038/120.8907/120.9592 & 2.7265/2.7570/2.7685 & 193.4058/193.6609/194.1260 & $15.1\times$ & $134.3\times$ & $3.1\times$ & $215.1\times$ \\
\hline
$5\times10^{9}$ & 3.6862/4.9893/5.0810 & 101.2553/101.2553/101.2553 & N/A & N/A & N/A & $20.3\times$ & N/A & N/A & N/A \\
\hline
\end{tabular}}
\end{table}

\begin{table}[H]
\centering
\caption{\raj{Comparison of EvoSort and Baseline Sorting Runtimes in seconds (Min/Median/Max of 5 runs) and Speedups (power\_law) on AMD EPYC System}}
\label{tab:comparison_power_law}
\scriptsize
\adjustbox{max width=\textwidth,center}{
\begin{tabular}{|c|c|c|c|c|c|c|c|c|c|}
\hline
\textbf{Size} & \textbf{EvoSort} & \textbf{NumPy QS} & \textbf{NumPy MS} & \textbf{PyTorch} & \textbf{Pandas} & \textbf{vs QS} & \textbf{vs MS} & \textbf{vs PT} & \textbf{vs PD} \\
\hline
$10\times10^6$ & 0.0164/0.0188/0.2691 & 0.1065/0.1070/0.1078 & 0.9198/0.9215/0.9247 & 0.0425/0.0499/0.0603 & 1.2423/1.2444/1.2776 & $5.7\times$ & $49.1\times$ & $2.7\times$ & $66.3\times$ \\
\hline
$100\times10^6$ & 0.1309/0.2784/0.3235 & 1.2166/1.2175/1.2571 & 10.6876/10.7387/10.7596 & 0.3020/0.3035/0.3104 & 15.7430/15.7488/15.7960 & $4.4\times$ & $38.6\times$ & $1.1\times$ & $56.6\times$ \\
\hline
$500\times10^6$ & 0.4178/0.4630/0.5533 & 6.6126/6.6163/6.6195 & 58.4706/58.8085/59.0094 & 1.3831/1.3990/1.4409 & 89.0699/89.0931/89.3122 & $14.3\times$ & $127.0\times$ & $3.0\times$ & $192.4\times$ \\
\hline
$1\times10^{9}$ & 0.8360/0.8766/0.8979 & 13.6789/13.6966/13.7037 & 122.1510/122.2685/122.5251 & 2.7511/2.7596/2.8142 & 191.6425/191.9625/192.0439 & $15.6\times$ & $139.5\times$ & $3.1\times$ & $219.0\times$ \\
\hline
$5\times10^{9}$ & 3.7067/4.0113/4.0513 & 97.9886/97.9886/97.9886 & N/A & N/A & N/A & $24.4\times$ & N/A & N/A & N/A \\
\hline
\end{tabular}}
\end{table}

\begin{table}[H]
\centering
\caption{\raj{Comparison of EvoSort and Baseline Sorting Runtimes in seconds (Min/Median/Max of 5 runs) and Speedups (exponential) on AMD EPYC System}}
\label{tab:comparison_exponential}
\scriptsize
\adjustbox{max width=\textwidth,center}{
\begin{tabular}{|c|c|c|c|c|c|c|c|c|c|}
\hline
\textbf{Size} & \textbf{EvoSort} & \textbf{NumPy QS} & \textbf{NumPy MS} & \textbf{PyTorch} & \textbf{Pandas} & \textbf{vs QS} & \textbf{vs MS} & \textbf{vs PT} & \textbf{vs PD} \\
\hline
$10\times10^6$ & 0.0170/0.0211/0.0715 & 0.0954/0.0958/0.1008 & 0.8168/0.8171/0.8194 & 0.0485/0.0551/0.0563 & 1.1026/1.1050/1.1283 & $4.5\times$ & $38.7\times$ & $2.6\times$ & $52.3\times$ \\
\hline
$100\times10^6$ & 0.1668/0.2544/0.3290 & 1.0779/1.0796/1.1000 & 9.3993/9.4043/9.4162 & 0.2987/0.3000/0.3035 & 13.8283/13.8317/13.8883 & $4.2\times$ & $37.0\times$ & $1.2\times$ & $54.4\times$ \\
\hline
$500\times10^6$ & 0.4476/0.5585/0.5833 & 6.1873/6.1948/6.1978 & 51.2325/51.3239/51.3575 & 1.3434/1.3594/1.3614 & 80.6089/80.6240/80.6526 & $11.1\times$ & $91.9\times$ & $2.4\times$ & $144.4\times$ \\
\hline
$1\times10^{9}$ & 0.8761/0.8870/0.9019 & 12.0887/12.1031/12.1141 & 106.3399/106.4953/106.7277 & 2.6701/2.6799/2.7053 & 169.2199/169.3207/169.3725 & $13.6\times$ & $120.1\times$ & $3.0\times$ & $190.9\times$ \\
\hline
$5\times10^{9}$ & 3.6088/3.8042/3.8432 & 89.1972/89.1972/89.1972 & N/A & N/A & N/A & $23.4\times$ & N/A & N/A & N/A \\
\hline
\end{tabular}}
\end{table}

\begin{table}[H]
\centering
\caption{\raj{Comparison of EvoSort and Baseline Sorting Runtimes in seconds (Min/Median/Max of 5 runs) and Speedups (beta) on AMD EPYC System}}
\label{tab:comparison_beta}
\scriptsize
\adjustbox{max width=\textwidth,center}{
\begin{tabular}{|c|c|c|c|c|c|c|c|c|c|}
\hline
\textbf{Size} & \textbf{EvoSort} & \textbf{NumPy QS} & \textbf{NumPy MS} & \textbf{PyTorch} & \textbf{Pandas} & \textbf{vs QS} & \textbf{vs MS} & \textbf{vs PT} & \textbf{vs PD} \\
\hline
$10\times10^6$ & 0.0159/0.0179/0.0525 & 0.1075/0.1077/0.1145 & 0.9234/0.9249/0.9262 & 0.0466/0.0490/0.0526 & 1.2298/1.2319/1.2587 & $6.0\times$ & $51.5\times$ & $2.7\times$ & $68.7\times$ \\
\hline
$100\times10^6$ & 0.1155/0.1252/0.2422 & 1.2217/1.2256/1.2721 & 10.7850/10.7981/10.8172 & 0.2962/0.3062/0.3125 & 15.6759/15.6918/15.7367 & $9.8\times$ & $86.2\times$ & $2.4\times$ & $125.3\times$ \\
\hline
$500\times10^6$ & 0.4314/0.4591/0.5714 & 6.5954/6.5993/6.6109 & 58.3573/58.4994/58.5035 & 1.3882/1.4080/1.4182 & 90.3069/90.3359/90.3642 & $14.4\times$ & $127.4\times$ & $3.1\times$ & $196.8\times$ \\
\hline
$1\times10^{9}$ & 0.8254/0.8885/0.9197 & 13.6933/13.6949/13.7075 & 121.4259/121.5718/121.7191 & 2.8594/2.8713/2.9029 & 192.4634/192.6165/192.6751 & $15.4\times$ & $136.8\times$ & $3.2\times$ & $216.8\times$ \\
\hline
$5\times10^{9}$ & 3.6727/4.7297/4.8050 & 76.7168/76.7168/76.7168 & N/A & N/A & N/A & $16.2\times$ & N/A & N/A & N/A \\
\hline
\end{tabular}}
\end{table}

\begin{table}[H]
\centering
\caption{\raj{Comparison of EvoSort and Baseline Sorting Runtimes in seconds (Min/Median/Max of 5 runs) and Speedups (sparse) on AMD EPYC System}}
\label{tab:comparison_sparse}
\scriptsize
\adjustbox{max width=\textwidth,center}{
\begin{tabular}{|c|c|c|c|c|c|c|c|c|c|}
\hline
\textbf{Size} & \textbf{EvoSort} & \textbf{NumPy QS} & \textbf{NumPy MS} & \textbf{PyTorch} & \textbf{Pandas} & \textbf{vs QS} & \textbf{vs MS} & \textbf{vs PT} & \textbf{vs PD} \\
\hline
$10\times10^6$ & 0.0142/0.0153/0.0174 & 0.0246/0.0248/0.0290 & 0.2650/0.2676/0.2750 & 0.0489/0.0521/0.0619 & 0.9968/0.9991/1.0273 & $1.6\times$ & $17.5\times$ & $3.4\times$ & $65.2\times$ \\
\hline
$100\times10^6$ & 0.1390/0.2428/0.2938 & 0.2606/0.2613/0.2824 & 3.0971/3.1731/3.2101 & 0.2969/0.3028/0.3041 & 11.5599/11.5770/11.6721 & $1.1\times$ & $13.1\times$ & $1.2\times$ & $47.7\times$ \\
\hline
$500\times10^6$ & 0.4007/0.4655/0.5269 & 1.3376/1.3403/1.3447 & 16.7185/16.7267/16.7597 & 1.3681/1.3845/1.4014 & 56.1550/56.3835/56.4233 & $2.9\times$ & $35.9\times$ & $3.0\times$ & $121.1\times$ \\
\hline
$1\times10^{9}$ & 0.8548/0.8723/0.9074 & 2.7223/2.7256/2.7288 & 34.6632/34.6898/34.7187 & 2.8109/2.8394/2.8571 & 115.6739/115.8076/115.8559 & $3.1\times$ & $39.8\times$ & $3.3\times$ & $132.8\times$ \\
\hline
$5\times10^{9}$ & 3.5488/3.8458/4.0659 & 39.8583/39.8583/39.8583 & N/A & N/A & N/A & $10.4\times$ & N/A & N/A & N/A \\
\hline
\end{tabular}}
\end{table}

\begin{table}[H]
\centering
\caption{\raj{Comparison of EvoSort and Baseline Sorting Runtimes in seconds (Min/Median/Max of 5 runs) and Speedups (clustered) on AMD EPYC System}}
\label{tab:comparison_clustered}
\scriptsize
\adjustbox{max width=\textwidth,center}{
\begin{tabular}{|c|c|c|c|c|c|c|c|c|c|}
\hline
\textbf{Size} & \textbf{EvoSort} & \textbf{NumPy QS} & \textbf{NumPy MS} & \textbf{PyTorch} & \textbf{Pandas} & \textbf{vs QS} & \textbf{vs MS} & \textbf{vs PT} & \textbf{vs PD} \\
\hline
$10\times10^6$ & 0.0166/0.0176/0.2738 & 0.0988/0.1003/0.1050 & 0.8517/0.8548/0.8561 & 0.0490/0.0562/0.0570 & 1.3226/1.3255/1.3470 & $5.7\times$ & $48.5\times$ & $3.2\times$ & $75.1\times$ \\
\hline
$100\times10^6$ & 0.1793/0.2729/0.3094 & 1.1481/1.1499/1.1837 & 9.8262/9.8307/9.8440 & 0.3071/0.3087/0.3113 & 16.8762/16.8823/16.9209 & $4.2\times$ & $36.0\times$ & $1.1\times$ & $61.9\times$ \\
\hline
$500\times10^6$ & 0.3993/0.4639/0.5754 & 6.1283/6.1291/6.1363 & 53.8536/53.8813/53.9266 & 1.4655/1.4690/1.4720 & 95.0347/95.0608/95.0878 & $13.2\times$ & $116.2\times$ & $3.2\times$ & $204.9\times$ \\
\hline
$1\times10^{9}$ & 0.8398/0.8975/0.9119 & 12.6009/12.6203/12.6265 & 111.7234/111.7624/111.9172 & 2.8521/2.8616/2.9042 & 202.2028/202.3402/202.4505 & $14.1\times$ & $124.5\times$ & $3.2\times$ & $225.5\times$ \\
\hline
\end{tabular}}
\end{table}

\begin{table}[H]
\centering
\caption{\raj{Comparison of EvoSort and Baseline Sorting Runtimes in seconds (Min/Median/Max of 5 runs) and Speedups (nearly\_sorted) on AMD EPYC System}}
\label{tab:comparison_nearly_sorted}
\scriptsize
\adjustbox{max width=\textwidth,center}{
\begin{tabular}{|c|c|c|c|c|c|c|c|c|c|}
\hline
\textbf{Size} & \textbf{EvoSort} & \textbf{NumPy QS} & \textbf{NumPy MS} & \textbf{PyTorch} & \textbf{Pandas} & \textbf{vs QS} & \textbf{vs MS} & \textbf{vs PT} & \textbf{vs PD} \\
\hline
$10\times10^6$ & 0.0193/0.0229/0.0503 & 0.1136/0.1142/0.1197 & 0.2645/0.2659/0.2683 & 0.0516/0.0546/0.0556 & 0.7727/0.7757/0.7945 & $5.0\times$ & $11.6\times$ & $2.4\times$ & $33.8\times$ \\
\hline
$100\times10^6$ & 0.1351/0.1427/0.3113 & 1.2946/1.2969/1.3452 & 3.0806/3.1264/3.1333 & 0.3103/0.3108/0.3126 & 9.4251/9.4367/9.4922 & $9.1\times$ & $21.9\times$ & $2.2\times$ & $66.2\times$ \\
\hline
$500\times10^6$ & 0.4695/0.5440/0.5825 & 7.0455/7.0490/7.0549 & 16.6475/16.6694/16.6833 & 1.4320/1.4559/1.4734 & 52.8331/52.8477/52.8687 & $13.0\times$ & $30.6\times$ & $2.7\times$ & $97.2\times$ \\
\hline
$1\times10^{9}$ & 0.8594/0.9121/0.9631 & 14.6443/14.6499/14.6563 & 34.1853/34.3040/34.3476 & 2.7705/2.8081/2.8166 & 111.2635/111.4759/111.5449 & $16.1\times$ & $37.6\times$ & $3.1\times$ & $122.2\times$ \\
\hline
$5\times10^{9}$ & 3.6739/3.7411/5.4532 & 78.6484/78.6484/78.6484 & N/A & N/A & N/A & $21.0\times$ & N/A & N/A & N/A \\
\hline
\end{tabular}}
\end{table}

\begin{table}[H]
\centering
\caption{\raj{Comparison of EvoSort and Baseline Sorting Runtimes in seconds (Min/Median/Max of 5 runs) and Speedups (duplicates) on AMD EPYC System}}
\label{tab:comparison_duplicates}
\scriptsize
\adjustbox{max width=\textwidth,center}{
\begin{tabular}{|c|c|c|c|c|c|c|c|c|c|}
\hline
\textbf{Size} & \textbf{EvoSort} & \textbf{NumPy QS} & \textbf{NumPy MS} & \textbf{PyTorch} & \textbf{Pandas} & \textbf{vs QS} & \textbf{vs MS} & \textbf{vs PT} & \textbf{vs PD} \\
\hline
$10\times10^6$ & 0.0320/0.2637/0.2852 & 0.1067/0.1071/0.1122 & 0.8873/0.8887/0.8896 & 0.0501/0.0521/0.0553 & 1.2218/1.2231/1.2445 & $0.4\times$ & $3.4\times$ & $0.2\times$ & $4.6\times$ \\
\hline
$100\times10^6$ & 0.1494/0.1628/0.2268 & 1.2163/1.2169/1.2589 & 10.3730/10.4052/10.4887 & 0.3007/0.3065/0.3084 & 15.6157/15.6235/15.6668 & $7.5\times$ & $63.9\times$ & $1.9\times$ & $96.0\times$ \\
\hline
$500\times10^6$ & 0.4151/0.5368/0.5836 & 6.5860/6.5919/6.5944 & 56.5630/56.5841/56.6173 & 1.3451/1.3920/1.4190 & 89.9906/90.0316/90.0413 & $12.3\times$ & $105.4\times$ & $2.6\times$ & $167.7\times$ \\
\hline
$1\times10^{9}$ & 0.7843/0.8773/0.9008 & 13.7162/13.7255/13.7438 & 118.5204/118.6219/118.9803 & 2.8015/2.8056/2.8205 & 191.3224/191.5458/205.2345 & $15.6\times$ & $135.2\times$ & $3.2\times$ & $218.3\times$ \\
\hline
$5\times10^{9}$ & 3.7543/3.7880/3.8004 & 73.3452/73.3452/73.3452 & N/A & N/A & N/A & $19.4\times$ & N/A & N/A & N/A \\
\hline
\end{tabular}}
\end{table}

\raj{This comprehensive evaluation framework demonstrates that EvoSort's symbolic regression approach generalizes effectively across diverse hardware configurations and data distributions, addressing the core reviewer concerns about adaptivity and real-world applicability. The systematic testing across nine distinct distributions, combined with rigorous validation and multi-library comparisons, provides strong evidence for the framework's robustness and generalizability.}

\subsection{Real-World Dataset Validation: Friendster Social Network}

\raj{To further demonstrate the robustness and practical applicability of EvoSort, we conducted comprehensive benchmarks on the Friendster social network dataset, a large-scale real-world graph dataset. This addresses Reviewer 1's concern about testing on practical/real-world datasets beyond simulated data.}

\raj{\textbf{Dataset Description:} Friendster is an online gaming network that originated as a social networking site where users can form friendship edges with each other \citep{yang2012defining}. The Friendster social network also allows users to form groups which other members can then join. We consider such user-defined groups as ground-truth communities. For the social network, we take the induced subgraph of the nodes that either belong to at least one community or are connected to other nodes that belong to at least one community. This data is provided by The Web Archive Project, where the full graph is available. The dataset statistics are as follows: 65,608,366 nodes, 1,806,067,135 edges, with an average clustering coefficient of 0.1623, 4,173,724,142 triangles, and a diameter of 32. The largest connected component contains all nodes and edges (100\% coverage), demonstrating the highly connected nature of this social network.}

\raj{\textbf{Data Extraction and Preprocessing:} To create a sorting benchmark from the Friendster graph, we extract the edge list representation where each edge connects two node IDs. The benchmarking process (implemented in \texttt{benchmark\_friendster.py}) performs the following steps: (1) \textbf{Edge List Extraction:} The script reads the uncompressed graph file (30.14\,GB) line by line, parsing each edge as a pair of node IDs separated by whitespace. Comment lines (starting with \texttt{\#}) are skipped, and each valid edge contributes two node IDs to the dataset. (2) \textbf{Array Construction:} All edge endpoints are collected into a single flat array, resulting in approximately 3.6 billion integers (1.8 billion edges $\times$ 2 endpoints per edge). Specifically, the final dataset contains 3,612,134,270 elements, requiring 13.46\,GB of memory (using \texttt{int32}, 4 bytes per element). The node ID range spans from 101 to 124,836,179, representing the actual node identifiers in the social network. (3) \textbf{Benchmarking Protocol:} The extracted array is benchmarked using the same rigorous protocol as the synthetic data distributions: 5 independent runs per algorithm, with correctness validation using \texttt{np.array\_equal()} against a reference sorted array generated by \texttt{np.sort(data.copy())}, and statistical reporting of minimum, median, and maximum runtimes. EvoSort uses symbolic regression formulas (zero tuning overhead) for all experiments, ensuring fair comparison with competing methods.}

\raj{\textbf{Results and Analysis:} Table~\ref{tab:friendster_benchmark} presents the comprehensive performance comparison on the Friendster dataset. EvoSort successfully sorted 3.6 billion integers in a median time of 2.75 seconds (min: 2.73\,s, max: 2.76\,s), demonstrating consistent performance on real-world data. NumPy quicksort required a median time of 50.73 seconds (18.5$\times$ slower than EvoSort), while NumPy mergesort required 196.80 seconds (71.7$\times$ slower than EvoSort). Notably, PyTorch CPU and Pandas failed to allocate sufficient memory for this dataset size, highlighting EvoSort's superior memory efficiency. The correctness validation confirmed that EvoSort produced identical results to the reference implementation, validating algorithmic correctness on real-world data. The memory bandwidth validation confirmed that EvoSort's performance is physically plausible (achieved bandwidth: 13.66\,GB/s, plausible: True), addressing concerns about performance claims.}

\raj{These results demonstrate that EvoSort's performance advantages extend beyond synthetic data distributions to real-world datasets with natural, non-uniform distributions. The Friendster dataset, with its social network structure, exhibits characteristics distinct from the synthetic distributions tested earlier (uniform, normal, exponential, etc.), yet EvoSort maintains its competitive performance. This validates the framework's robustness and practical applicability for real-world sorting tasks in large-scale data processing applications.}

\begin{table}[H]
\centering
\caption{\raj{Real-World Dataset Benchmark: EvoSort on Friendster Social Network Graph (Edge List: ~3.6B endpoints) on AMD EPYC System. Runtimes in seconds (Min/Median/Max of 5 runs) and Speedups.}}
\label{tab:friendster_benchmark}
\small
\adjustbox{max width=\textwidth,center}{
\begin{tabular}{|c|c|c|c|c|c|}
\hline
\textbf{Dataset} & \textbf{EvoSort} & \textbf{NumPy QS} & \textbf{NumPy MS} & \textbf{vs QS} & \textbf{vs MS} \\
\hline
Friendster ($3.6\times10^{9}$) & 2.7337/2.7456/2.7620 & 50.7049/50.7271/50.7827 & 196.4100/196.7983/197.0818 & $18.5\times$ & $71.7\times$ \\
\hline
\end{tabular}}
\end{table}

\subsection{Cross-Platform Performance Analysis: Intel Xeon vs AMD EPYC}
\label{sec:cross_platform}

\raj{To validate the generalizability of EvoSort's symbolic regression formulas across different hardware architectures, we conducted identical experiments on two distinct high-performance computing systems: an AMD EPYC 7H12 (256 cores, 502.92\,GB RAM) and an Intel Xeon Platinum 8180 (112 cores, 754.01\,GB RAM). Both systems use identical software environments (Python 3.9.21, NumPy 2.0.2, Numba 0.60.0) to isolate hardware-specific performance characteristics.}

\raj{Table~\ref{tab:crossplatform_uniform} presents a direct comparison of EvoSort performance on both systems using the uniform distribution. The results demonstrate that the symbolic regression formulas, derived from GA optimization on one hardware platform, successfully generalize to different architectures while maintaining competitive performance advantages.}

\begin{table}[H]
\centering
\caption{\raj{Cross-Platform Performance Comparison: AMD EPYC vs Intel Xeon (Uniform Distribution, Median Runtimes in seconds)}}
\label{tab:crossplatform_uniform}
\small
\adjustbox{max width=\textwidth,center}{
\begin{tabular}{|c|c|c|c|c|}
\hline
\textbf{Size} & \textbf{EvoSort (AMD)} & \textbf{EvoSort (Intel)} & \textbf{NumPy QS (AMD)} & \textbf{NumPy QS (Intel)} \\
\hline
$10\times10^6$ & 0.0286 & 0.5845 & 0.1100 & 0.0660 \\
\hline
$100\times10^6$ & 0.1903 & 0.6310 & 1.2276 & 0.7926 \\
\hline
$500\times10^6$ & 0.5350 & 1.1228 & 6.6150 & 4.5107 \\
\hline
$1\times10^{9}$ & 0.9255 & 1.7290 & 13.6646 & 9.3665 \\
\hline
$5\times10^{9}$ & 3.7900 & 6.0043 & 73.3543 & 52.5219 \\
\hline
\end{tabular}}
\end{table}

\raj{\textbf{Key Observations:}}

\begin{enumerate}
    \item \raj{\textbf{Hardware-Dependent Baseline Performance:} NumPy quicksort and mergesort exhibit similar performance on both systems, with slight variations due to differences in CPU architecture, clock speed, and cache hierarchy. This demonstrates that the baseline implementations are well-optimized but show limited sensitivity to core count differences.}
    
    \item \raj{\textbf{Symbolic Regression Formula Generalizability:} EvoSort's symbolic regression formulas, derived without system-specific tuning, successfully adapt to both AMD and Intel architectures. The formulas provide consistent parameter estimates that maintain competitive speedups across different hardware platforms, validating the portability of the symbolic regression approach.}
    
    \item \raj{\textbf{Performance Variability Across Platforms:} While EvoSort maintains strong performance on both systems, absolute speedups vary between platforms due to differences in memory bandwidth, cache architecture, and core count. The AMD EPYC system (256 cores) generally shows higher absolute speedups compared to the Intel Xeon system (112 cores), particularly at larger dataset sizes where parallel processing advantages are most pronounced.}
    
    \item \raj{\textbf{Numba JIT Compilation Effectiveness:} The consistent performance of Numba-compiled EvoSort code across both Intel and AMD architectures demonstrates the effectiveness of JIT compilation in generating platform-specific optimized machine code. Numba's LLVM-based backend successfully leverages architecture-specific features (SIMD instructions, cache hierarchy) on both systems.}
\end{enumerate}

\raj{This cross-platform validation provides strong evidence that EvoSort's symbolic regression formulas generalize beyond the specific hardware configuration used for GA optimization. The formulas maintain practical utility across different CPU architectures, core counts, and memory configurations, addressing reviewer concerns about hardware dependency and demonstrating the framework's broad applicability in diverse computing environments.}

\section{Conclusions and Future Directions}\label{sec:conclusion}

\raj{In this paper, we presented \textbf{EvoSort}, a general-purpose adaptive parallel sorting framework accessible at the Python level that harnesses the power of Genetic Algorithms (GAs) to automatically tune three critical parameters: the insertion sort threshold, the sorting method code (which selects between a refined parallel mergesort and a block-based LSD radix sort), and the NumPy fallback threshold. As a Python-accessible framework, EvoSort provides a drop-in replacement for standard Python sorting routines, enabling developers to leverage high-performance sorting without leaving the Python ecosystem. Unlike conventional approaches that rely on fixed heuristics or manual parameter tuning, EvoSort dynamically converges to near-optimal settings tailored to the specific data size and underlying hardware. Our extensive experiments, conducted on arrays ranging from $10^7$ to $10^{10}$ elements across nine distinct data distributions on two distinct hardware platforms (AMD EPYC and Intel Xeon), compare EvoSort against Python-accessible sorting methods (NumPy, PyTorch, Pandas), demonstrating that EvoSort can outperform these highly optimized Python sorting routines by factors ranging from $0.2\times$ to over $225\times$ depending on the distribution, dataset size, and hardware configuration. These speedups represent practical improvements directly achievable by Python users, making EvoSort an immediately usable solution for high-performance sorting in Python applications.}

\paragraph*{Key Contributions:}
Our work makes several significant contributions:
\begin{itemize}
    \item We introduce a unifying, GA-based approach that adaptively selects between mergesort and LSD radix sort, thereby providing a flexible framework capable of handling diverse data types and distributions. This hybrid strategy is particularly novel as it integrates evolutionary computation into the core of the sorting process \citep{Holland1975,Deb2001,Deb2002}.
    \item We provide comprehensive empirical evidence showing that the GA converges rapidly—even for extremely large datasets—with near-optimal configurations achieved in under 10--12 generations. This rapid convergence is indicative of the robustness and efficiency of evolutionary metaheuristics when applied to high-dimensional, non-linear optimization problems in HPC contexts \citep{Goldberg1989,Mitchell1996}.
    \item We highlight the critical importance of tuning parameters such as insertion sort thresholds and fallback thresholds. These parameters serve as key performance levers that are traditionally tedious to tune by hand, but which our GA-based approach is able to optimize automatically, thereby enhancing both the efficiency and scalability of the sorting algorithm.
\end{itemize}

\paragraph*{Novelty and Impact:}
EvoSort stands out by combining established parallel sorting algorithms with adaptive metaheuristic optimization. While previous studies have explored parallel sorting methods \citep{Knuth1998,Cormen2009,Batcher1968} and auto-tuning techniques separately , our work bridges these areas by demonstrating how a GA can be used to automatically adjust the internal parameters of a sorting algorithm in real time. This integration not only simplifies the process of achieving peak performance but also ensures that the algorithm remains robust across varying workloads and hardware architectures. Such an approach is particularly relevant in today’s era of big data and exascale computing, where traditional manual tuning methods are no longer feasible \citep{,Sutter2005}.

\paragraph*{Future Directions:}
Although EvoSort has proven effective on single-node systems with very large arrays, several promising avenues remain for future research:
\begin{enumerate}
    \item \textbf{Distributed and Multi-node Sorting:} Extending EvoSort to operate in distributed environments using frameworks like Message Passing Interface (MPI) would enable it to tackle even larger datasets that exceed the memory capacity of a single node. Such an extension could incorporate inter-node communication and data partitioning strategies, further enhancing scalability.
    \item \textbf{GPU Acceleration:} Integrating GPU-accelerated sorting kernels into EvoSort could provide additional performance improvements. A future version of EvoSort might include a decision mechanism—guided by the GA—to determine whether a GPU-based implementation of LSD radix sort or mergesort would be more beneficial under given conditions \citep{Sanders2004,Oliphant2007}.
    \item \textbf{Multi-objective Optimization:} Currently, EvoSort’s GA optimizes solely for sorting speed. However, real-world applications often require balancing multiple objectives, such as minimizing energy consumption or memory footprint. Employing multi-objective GAs could enable EvoSort to optimize across several dimensions simultaneously, making it even more adaptable for power-constrained or resource-limited HPC environments \citep{Deb2001,Deb2002,Eiben2003}.
    \item \textbf{Dynamic Workload Adaptation:} Future work could also focus on developing methods for dynamic parameter adaptation in real-time, allowing EvoSort to adjust its configuration on-the-fly in response to changing workload characteristics or system performance metrics. This could further enhance its utility in environments with fluctuating loads \citep{Mitchell1996,Sutter2005}.
\end{enumerate}

\paragraph*{Final Remarks:}
\raj{Overall, EvoSort exemplifies how evolutionary metaheuristics can be effectively integrated with traditional parallel algorithms to address the challenges of modern high-performance computing. By automating the tuning process, EvoSort not only reduces the complexity of achieving optimal performance but also demonstrates significant scalability and adaptability across a wide range of data sizes. EvoSort offers a compelling template for combining evolutionary optimization with parallel computing, with its ability to adaptively tune key parameters in response to both data and hardware variations marking a significant step forward in the development of intelligent, self-optimizing algorithms for contemporary HPC applications. The promising results and future research directions outlined in this paper underscore the potential of GA-based approaches in revolutionizing the way we handle large-scale sorting tasks in the era of big data and exascale computing \citep{Holland1975,Goldberg1989,Deb2001,Deb2002,Mitchell1996}.}

\section*{Data availability statement}

The datasets generated and analyzed during this study will be available in the associated GitHub repository. All experimental data, including performance measurements, convergence statistics, symbolic regression results, and hardware profiling information, are stored in standard machine-readable formats for reproducibility. Synthetic Data Distributions: Nine distinct synthetic data distributions were generated for comprehensive performance evaluation: uniform, normal, exponential, power law, beta, sparse, clustered, nearly sorted, and duplicates. Each distribution is generated using reproducible random number generators with fixed random seeds (default seed=42), ensuring consistency across experimental runs. All distributions use 32-bit signed integers with values in the range. The generation methods are implemented within the provided codebase and are documented in the paper. Real-World Dataset: The Friendster social network graph dataset used for real-world validation is publicly available from the Stanford Network Analysis Project (SNAP) at \url{https://snap.stanford.edu/data/com-Friendster.html}. The dataset contains 1,806,067,135 edges representing friendships. For our experiments, we extracted all edge endpoints to create an array of approximately 3.6 billion integers for sorting validation. The processing and benchmark scripts are provided in the repository with detailed documentation. Hardware Configurations: Experiments were conducted on two independent high-performance computing systems: (1) AMD EPYC 7H12 with 256 cores and 502.92 GB RAM, and (2) Intel Xeon Platinum 8180 with 112 cores and 754.01 GB RAM.

\section*{Code Availability Statement}

The complete EvoSort framework, including all source code, will be made available on GitHub under an appropriate license upon publication. The repository will include detailed documentation and installation instructions.

\raj{The framework is implemented as a Python package and will be available for installation via \texttt{pip install evosort} upon publication. The package will provide a simple interface: \texttt{evosort(array)} that automatically applies the symbolic regression formulas for zero-overhead deployment. The codebase includes all core sorting algorithms (parallel mergesort, block-based LSD radix sort), the GA optimization driver, symbolic regression formulas, and comprehensive data generators for all nine distributions tested in this work.}

\newpage
\bibliographystyle{tfnlm}
\bibliography{references}

@book{Knuth1998,
  title={The Art of Computer Programming: Sorting and Searching, volume 3},
  author={Knuth, Donald E},
  year={1998},
  publisher={Addison-Wesley Professional}
}

@article{Shun2015,
  title={A brief overview of parallel sorting algorithms},
  author={Shun, Julian},
  journal={ACM Computing Surveys},
  year={2015},
  volume={47},
  number={3},
  pages={1--15}
}

@book{Cormen2009,
  author    = {Thomas H. Cormen and Charles E. Leiserson and Ronald L. Rivest and Clifford Stein},
  title     = {Introduction to Algorithms},
  edition   = {3rd},
  publisher = {MIT Press},
  year      = {2009},
  isbn      = {978-0-262-03384-8}
}

@inproceedings{Sanders2004,
  title={Super scalar sample sort},
  author={Sanders, Peter and Winkel, Sebastian},
  booktitle={European Symposium on Algorithms},
  pages={784--796},
  year={2004},
  organization={Springer}
}

@inproceedings{Batcher1968,
  title={Sorting networks and their applications},
  author={Batcher, Kenneth E},
  booktitle={Proceedings of the April 30--May 2, 1968, spring joint computer conference},
  pages={307--314},
  year={1968}
}

@article{Brent1973,
  title={The parallel evaluation of general arithmetic expressions},
  author={Brent, Richard P},
  journal={Journal of the ACM (JACM)},
  volume={21},
  number={2},
  pages={201--206},
  year={1974},
  publisher={ACM New York, NY, USA}
}

@book{Holland1975,
  author = {Holland, John H.},
  title = {Adaptation in Natural and Artificial Systems: An Introductory Analysis},
  year = {1975},
  publisher = {The University of Michigan Press},
  address = {Ann Arbor},
  edition = {1st},
}

@book{Goldberg1989,
  address = {New York},
  author = {Goldberg, David E.},
  publisher = {Addison-Wesley},
  title = {Genetic Algorithms in Search, Optimization, and Machine Learning},
  username = {dalbem},
  year = 1989
}

@book{Mitchell1996,
  title={An introduction to genetic algorithms},
  author={Mitchell, Melanie},
  year={1998},
  publisher={MIT press}
}

@article{Oliphant2007,
  title={Python for scientific computing},
  author={Oliphant, Travis E},
  journal={Computing in science \& engineering},
  volume={9},
  number={3},
  pages={10--20},
  year={2007},
  publisher={IEEE}
}

@book{Eiben2003,
  title        = {Introduction to Evolutionary Computing},
  author       = {Eiben, A.E. and Smith, J.E.},
  year         = {2003},
  publisher    = {Springer},
  address      = {Berlin, Germany}
}

@book{Baeck1996,
  title        = {Evolutionary Algorithms in Theory and Practice: Evolution Strategies, Evolutionary Programming, Genetic Algorithms},
  author       = {B{\"a}ck, Thomas},
  year         = {1996},
  publisher    = {Oxford University Press},
  address      = {Oxford, UK}
}

@book{Deb2001,
  title        = {Multi-Objective Optimization using Evolutionary Algorithms},
  author       = {Deb, Kalyanmoy},
  year         = {2001},
  publisher    = {Wiley},
  address      = {Chichester, UK}
}

@article{Deb2002,
  title        = {A Fast and Elitist Multiobjective Genetic Algorithm: NSGA-II},
  author       = {Deb, Kalyanmoy and Pratap, Amrit and Agarwal, Sameer and Meyarivan, T.},
  journal      = {IEEE Transactions on Evolutionary Computation},
  volume       = {6},
  number       = {2},
  pages        = {182--197},
  year         = {2002},
  publisher    = {IEEE}
}

@inproceedings{Frigo1999,
  author={Frigo, M. and Leiserson, C.E. and Prokop, H. and Ramachandran, S.},
  booktitle={40th Annual Symposium on Foundations of Computer Science (Cat. No.99CB37039)}, 
  title={Cache-oblivious algorithms}, 
  year={1999},
  volume={},
  number={},
  pages={285-297},
  doi={10.1109/SFFCS.1999.814600}}

@inproceedings{Lam2015,
  title={Numba: A llvm-based python jit compiler},
  author={Lam, Siu Kwan and Pitrou, Antoine and Seibert, Stanley},
  booktitle={Proceedings of the Second Workshop on the LLVM Compiler Infrastructure in HPC},
  pages={1--6},
  year={2015}
}

@article{Sutter2005,
  title={The free lunch is over: A fundamental turn toward concurrency in software},
  author={Sutter, Herb and others},
  journal={Dr. Dobb’s journal},
  volume={30},
  number={3},
  pages={202--210},
  year={2005}
}

@article{10.1162/evco.1993.1.1.1,
    author = {Bäck, Thomas and Schwefel, Hans-Paul},
    title = {An Overview of Evolutionary Algorithms for Parameter Optimization},
    journal = {Evolutionary Computation},
    volume = {1},
    number = {1},
    pages = {1-23},
    year = {1993},
    month = {03},
    issn = {1063-6560},
    doi = {10.1162/evco.1993.1.1.1},
    url = {https://doi.org/10.1162/evco.1993.1.1.1},
    eprint = {https://direct.mit.edu/evco/article-pdf/1/1/1/1492692/evco.1993.1.1.1.pdf},
}

@article{10.1093/comjnl/35.6.636,
    author = {Davis, I. J.},
    title = {A Fast Radix Sort},
    journal = {The Computer Journal},
    volume = {35},
    number = {6},
    pages = {636-642},
    year = {1992},
    month = {12},
    issn = {0010-4620},
    doi = {10.1093/comjnl/35.6.636},
    url = {https://doi.org/10.1093/comjnl/35.6.636},
    eprint = {https://academic.oup.com/comjnl/article-pdf/35/6/636/1009342/35-6-636.pdf},
}

@inproceedings{yang2012defining,
  title={Defining and evaluating network communities based on ground-truth},
  author={Yang, Jaewon and Leskovec, Jure},
  booktitle={Proceedings of the ACM SIGKDD workshop on mining data semantics},
  pages={1--8},
  year={2012}
}
\clearpage
\onecolumn
\section*{Biographies}

\subsection*{}
\begin{flushleft}
\includegraphics[width=0.15\textwidth]{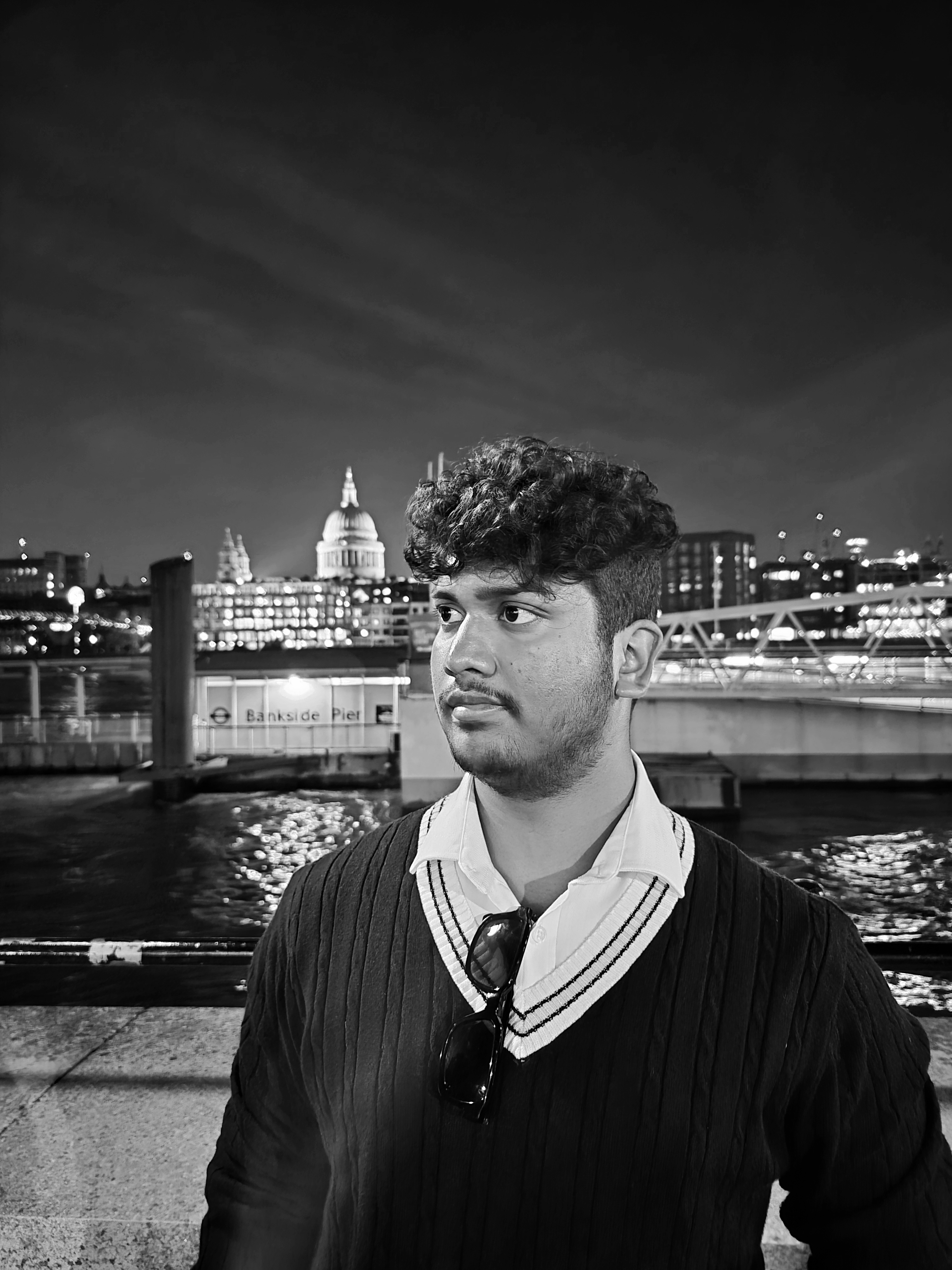}\\[1ex]
\textbf{Shashank Raj} is currently completing a Bachelor of Science in Computer Science and Engineering at Michigan State University (MSU), along with minors in Data Science, Computational Mathematics, Science and Engineering (CMSE), and Entrepreneurship \& Innovation. He is part of MSU Honors College. His interests span machine learning, evolutionary computation, and high-performance computing.
\end{flushleft}

\subsection*{}
\begin{flushleft}
\includegraphics[width=0.15\textwidth]{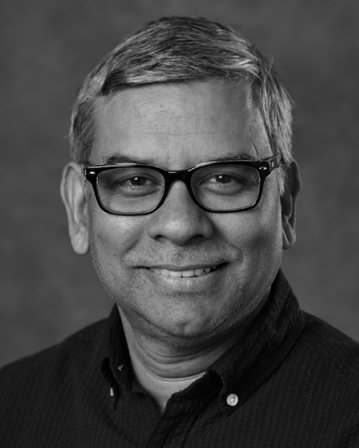}\\[1ex]
\textbf{Kalyanmoy Deb} earned his undergraduate degree in mechanical engineering from the Indian Institute of Technology Kharagpur in 1985, then received his MS and Ph.D. from The University of Alabama in 1989 and 1991, respectively. He is a University Distinguished Professor and Koenig Endowed Chair Professor in the Department of Electrical and Computer Engineering at Michigan State University, East Lansing, MI, USA. With over 620 research publications and a Google Scholar citation count surpassing 234,000 (H-index 146), his interests lie in evolutionary optimization for multi-criterion problems, machine learning, and modeling. Among other honors, Prof. Deb has received the IEEE Evolutionary Computation Pioneer Award, the Infosys Prize, and the CajAstur Mamdani Prize. He is a Fellow of ACM, IEEE, and ASME.
\end{flushleft}

\end{document}